\documentclass[authoryear,1p]{elsarticle}
\usepackage{graphicx}
\usepackage{array}
\usepackage[table]{xcolor}
\usepackage{multirow}
\usepackage{amssymb}
\usepackage{booktabs}
\usepackage{makecell}
\usepackage{empheq}
\usepackage{natbib}
\usepackage{appendix}
\usepackage{subfig}
\usepackage{xcolor}
\usepackage{eurosym}

\usepackage{url}
\usepackage{float}
\usepackage[T1]{fontenc}
\usepackage[utf8]{inputenc}
\usepackage{lmodern}
\usepackage{pdflscape}
\usepackage{tikz}
\definecolor{darkgreen}{rgb}{0.0, 0.5, 0.0}
\usepackage[colorinlistoftodos, textsize=tiny]{todonotes}
\usepackage[addedmarkup=uline, deletedmarkup=sout, authormarkup=none]{changes}
\definechangesauthor[name={authorname}, color=darkgreen]{add}
\definechangesauthor[name={authorname}, color=red]{del}

\usetikzlibrary{arrows.meta, positioning, fit, backgrounds}

\tikzset{
    arrowstyle/.style={draw=black, thick, -Latex}
}
\usetikzlibrary{positioning, arrows.meta, shapes}

\usepackage{enumitem}
\usepackage{nomencl}
\makenomenclature

\usepackage{mathtools}

\newdefinition{rmk}{Remark}

\usepackage{hyperref}
\newcommand{\sref}[2]{\hyperref[#2]{#1 \ref*{#2}}}

\let\oldref\ref
\renewcommand{\ref}[1]{(\oldref{#1})}
\usepackage[nameinlink, capitalise, noabbrev]{cleveref}
\hypersetup{
	colorlinks,
	citecolor=blue,
	filecolor=blue,
	linkcolor=blue,
	urlcolor=blue}
\crefname{prop}{Proposition}{Propositions}
\crefname{thm}{Theorem}{Theorems}
\crefname{cor}{Corollary}{Corollaries}
\crefname{ass}{Assumption}{Assumptions}
\makenomenclature
\usepackage{etoolbox}
\makeatletter
\def\bs{\expandafter\@gobble\string\\}
\def\lb{\expandafter\@gobble\string\{}
\def\rb{\expandafter\@gobble\string\}}
\def\@pdfauthor{C.V.Radhakrishnan}
\def\@pdftitle{elsarticle.cls -- A documentation}
\def\@pdfsubject{Document formatting with elsarticle.cls}
\def\@pdfkeywords{LaTeX, Elsevier Ltd, document class}

\DeclareRobustCommand{\LaTeX}{L\kern-.26em%
	{\sbox\z@ T%
		\vbox to\ht\z@{\hbox{\check@mathfonts
				\fontsize\sf@size\z@
				\math@fontsfalse\selectfont
				A\,}%
			\vss}%
	}%
	\kern-.15em%
	\TeX}
\journal{a Journal}

\begin{document}
	\begin{frontmatter}
		\title{Dynamic tariff-based demand response in retail electricity market under uncertainty}
		\author[1]{Arega Getaneh Abate}\corref{cor1}
		\ead{ageab@dtu.dk}
		\author[2]{Rossana Riccardi}
		\ead{rossana.riccardi@unibs.it}
		\author[3,4]{Carlos Ruiz}
		\ead{caruizm@est-econ.uc3m.es}
		\cortext[cor1]{Corresponding author}
		\address[1]{DTU Wind and Energy Systems, Technical University of Denmark, Kgs. Lyngby, Denmark}
		\address[2]{Department of Economics and Management,
			University of Brescia, 74/B Brescia, Italy}
		\address[3]{Department of Statistics, University Carlos III de Madrid, Avda. de la Universidad 30, 28911-Leganés, Spain}
		\address[4]{UC3M-BS Institute for Financial Big Data (FiBiD), Universidad Carlos III de Madrid, 28903, Getafe, Madrid, Spain}

				\begin{abstract}

Demand response (DR) programs play a crucial role in improving system reliability and mitigating price volatility by altering the core profile of electricity consumption. This paper proposes a game-theoretical model that captures the dynamic interplay between retailers (leaders) and consumers (followers) in a tariff-based electricity market under uncertainty. The proposed procedure offers theoretical and economic insights by analyzing consumer flexibility within a hierarchical decision-making framework. In particular, two main market configurations are examined under uncertainty: \textit{i}) there exists a retailer that exercises market power over consumers, and \textit{ii}) the retailer and the consumers participate in a perfect competitive game. The former is formulated as a mathematical program with equilibrium constraints (MPEC), whereas the latter is recast as a mixed-integer linear program (MILP). These problems are solved by deriving equivalent tractable reformulations based on the Karush-Kuhn-Tucker (KKT) optimality conditions of each agent's problem.

Numerical simulations based on real data from the European Energy Exchange platform are used to illustrate the performance of the proposed methodology. The results indicate that the proposed model effectively characterizes the interactions between retailers and flexible consumers in both perfect and imperfect market structures. Under perfect competition, the economic benefits extend not only to consumers but also to overall social welfare. Conversely, in an imperfect market, retailers leverage consumer flexibility to enhance their expected profits, transferring the risk of uncertainty to end-users. Additionally, the degree of consumer flexibility and consumers' valuation of electricity consumption play significant roles in shaping market outcomes. These findings highlight the crucial impact of market structure and consumer behavior on the dynamics of electricity market pricing under demand response programs.
		\end{abstract}
	\end{frontmatter}
 
\section{Introduction}

 The integration of renewable energy sources (RES) is rapidly increasing, driven by their environmentally friendly properties and economic benefits. Key global players such as the European Union, the United States, and China have set ambitious RES generation targets, aiming for 100\%, 80\%, and 60\%, respectively, by the year 2050. This green transition aims at reducing the reliance on emitting power sources, thereby decreasing CO$_2$ emissions and advancing towards achieving net-zero carbon goals. Despite these benefits, high RES penetration introduces significant uncertainty and system imbalances. These factors present considerable challenges to maintaining the reliability and security of power systems. Hence, it is of key importance to determine optimal operation strategies in electricity markets that ensure the real-time balance of supply and demand, considering the high penetration of RES and demand uncertainties \citep{he2013engage,zugno2013bilevel,niromandfam2020modeling, abate2022contract}. For instance, most of the markets where electricity is traded are contracted and cleared weeks, days, or hours before the energy delivery takes place and uncertainty is resolved. In such cases, the market may impose penalties or incur high curtailment costs if the dispatched power is not consumed as planned in real-time. Moreover, this imbalance can compromise system stability and requires the use of fast-response conventional units.
 Therefore, to enable the large-scale integration of RES and to enhance the decarbonization of electricity systems without endangering the security of supply, additional flexibility must be provided in the form of demand-side flexibility \citep{wang2020economic,morales2021reducing}. 
 
One potential solution is demand response (DR), which is typically achieved through incentive-based or price-based consumer flexibility. It works by offering consumers financial incentives to reduce or shift their peak loads to off-peak periods. To that end, consumers facing a high price during a given hour can adopt one of the following strategies (or a combination of them): \textit{i}) reducing consumption, \textit{ii}) shifting consumption to low-price periods, and/or \textit{iii}) covering part of the consumption by on-site distributive generation or storage sources. This behavior can be used to soften peak loads at given hours and reduce electricity market prices caused by marginal carbon-based technologies \citep{nolan2015challenges,guo2022dynamic}. Moreover, since DR is a distributed resource located at the end of the distribution system, additional environmental benefits are obtained from a reduction in electricity losses in the transmission and distribution lines \citep{shen2014role}. Thus, the main advantages of DR programs are: \textit{i}) they can contribute to reducing system costs and managing price volatility in real-time \citep{zugno2013bilevel, aussel2020trilevel, alipour2019real}; \textit{ii}) they allow higher penetration of intermittent RES in the electric power system  \citep{morales2013integrating,hakimi2019optimal,antunes2020bilevel}; \textit{iii}) they help improve system security by contributing to the real-time balance between generation and demand \citep{zugno2013bilevel,wang2020economic}; and \textit{iv}) they can mitigate electricity retailers' market power by increasing consumer flexibility \citep{devine2023role}. 
 
 Upon receiving updated information regarding price tariffs, consumers can calculate their benefits and the trade-off of shifting their consumption at critical times for system operators. The literature on DR extensively explores various incentive-based strategies. These include offering consumers dynamic electricity tariffs, such as time-of-use (TOU), critical peak pricing (CPP), and real-time pricing (RTP).  Dynamic tariffs can be attractive to retailers because they allow them to align retail price tariffs with spot market prices, thus transferring some of the retailer-centered risks to customers \citep{guo2022dynamic,askeland2020stochastic,clastres2021dynamic}. 
  Using tariff-based DR, \cite{askeland2020stochastic} formulate an electricity grid tariff design problem with bilevel programming in the context of prosumers at the end-user level.  They address the tariff optimization problem as a tool for indirect load control. By developing mathematical problems with equilibrium constraints (MPECs), they analyze how a simple tariff scheme can enhance prosumers' flexibility and efficiently reduce grid load. Applying a bilevel programming problem, \cite{clastres2021dynamic} study the impact of dynamic pricing and load-shifting on a retailer's electricity supply to consumer markets.  In a related study, \cite{guo2022dynamic} model a dynamic price-based DR program, incorporating the consumer benefit function. They find that implementing demand-side management incentives can reduce consumer electricity demand during peak hours and increase retailer profits.

However, significant gaps persist in the literature as the integration of RES grows and the demand for consumer flexibility increases. Key areas include: \textit{i}) incorporating the characterization of uncertainties affecting electricity prices, production costs, and demand in the day-ahead market; \textit{ii}) explicitly studying consumer behavior and leveraging flexibility for operational and planning decisions; and \textit{iii}) comparing perfect and imperfect electricity markets, involving multiple participants with varying market powers to analyze both theoretical and practical economic implications.

 Overall, the increase in energy costs, the global competition for fuel, and the tightening of emission requirements place considerable pressure on power systems. This motivates us to extend the potential of DR programs to enhance system reliability, ensure supply security, mitigate market power, and manage price volatility. These benefits are achieved while accounting for uncertainty across various market configurations.
 
 The objective of this work is to model and analyze the decision-making process and potential market interactions between electricity retailers and flexible consumers participating in the day-ahead electricity market. Specifically, we seek to mimic the hierarchy of the decision-making process, where a retailer needs to first set electricity price tariffs and the consumers react accordingly. Moreover, motivated by the concept of \textit{smart grids, smart devices}, such as batteries, EVs, or heat pumps, which promote different forms of horizontal consumer aggregations (cooperatives, energy communities, virtual power plants, etc.), we assume two market configurations. In the first one, the retailer has market power by anticipating the reactions of consumers, which is modeled by a bilevel (Stackelberg) problem. In the second one, we assume that the retailer is myopic to the consumers' response, which is modeled with a competitive equilibrium. Furthermore, for both configurations and extending previous works, we model the characterization of the consumers' flexibility by using a (nonlinear) quadratic utility function. 

These factors, along with varying levels of market competition and sources of uncertainty, influence retailers' and consumers' decision-making processes. This, in turn, impacts the resulting market outcomes. Therefore, this work models the profit maximization problems of retailers and the utility maximization problems of consumers in electricity consumption, taking into account the inherent uncertainties present in both perfect and imperfect electricity markets.  Specifically, the retailer's problem with market power is formulated as a bilevel program and solved \textit{as an MPEC}. For a comprehensive treatment of bilevel programs and their solution methods, the interested reader is referred to \cite{luo1996mathematical}. The equilibrium model with perfect competition, on the other hand, is solved via mixed-integer linear programming (MILP). We model uncertainty with stochastic programming, which extends the majority of deterministic DR models available in the literature. 

This paper presents several key findings: \textit{i}) It demonstrates that retailers can maximize their expected profits by exercising market power and utilizing consumer flexibility to mitigate/transfer their risk. \textit{ii}) The results show that consumers can minimize their electricity consumption costs and increase overall social welfare under a perfectly competitive market.  \textit{iii}) The proposed models can be adapted to various groups of consumers with flexibility. They can also include prosumers who trade electricity, either jointly or in equilibrium, to achieve individual objectives across different market configurations. 
  \textit{iv}) Explicitly modeling consumer behavior through a utility function in a dynamic pricing environment offers managerial insights. This approach helps to characterize and leverage consumer flexibility to optimize social welfare, especially with high RES penetration. Another takeaway is the increased vulnerability of the electricity market to the market power of oligopolistic entities when consumer flexibility is neglected, leading to market inefficiencies. Harnessing consumer flexibility in operational and strategic planning decisions with a competitive market structure can counterbalance retailers’ market power and mitigate grid instability.
 
 The remainder of this paper is organized as follows. \cref{SEC1} reviews the current literature on this topic. \cref{se_P_3_1} demonstrates the model formulation, which entails the retailer's and consumers' problems. The formulation of the problems as an MPEC and the linear reformulation of the problem as an MILP in equilibrium is presented. The computational and simulation results are reported and discussed in \cref{se_P_3_2}. Finally, \cref{se_P_3_3} closes by drawing some concluding remarks from the provided discussions and results. 

 \section{Literature review}\label{SEC1} 
  
 On the supply side, power producers, both conventional and RES, sell their generated power to the grid through organized energy markets. These markets are responsible for handling transactions related to energy and ancillary services. On the demand side, DR programs play an integral role in balancing and operational security in power systems. These programs, managed by DR buyers such as load-serving entities, distribution system operators (DSOs), large consumers, and DR aggregators, are pivotal in implementing strategies to collect and manage flexibility options from end-users. This management effectively modifies energy consumption patterns in response to supply conditions or market signals.

DR buyers primarily focus on interacting with consumers through contracts to create system flexibility. DR programs are crucial in day-ahead and intra-day markets, where supply and demand adjustments occur closer to real-time operations. The specific nature and extent of these interactions and transactions can vary, depending on the particular market design and regulatory environment.

In the literature, DR programs have been considered to address various economic and operational challenges. These challenges include achieving economic efficiency, reducing market volatility, optimizing utility in electricity consumption, minimizing discomfort in existing consumption patterns, enhancing grid stability, optimizing system performance, and reducing the carbon footprint.  \cite{devine2023role}, \cite{guo2022dynamic}, \cite{askeland2020stochastic}, \cite{clastres2021dynamic} and\cite{nilsson2018household}, among others, have explored and proposed models to address some of these challenges.
From retailers' perspective, dynamic tariff-based DR can be used in spot markets to transfer some of retailers' risks to consumers. As highlighted by \cite{nilsson2018household}, dynamic tariffs are recognized as one of the most effective strategies for enhancing demand flexibility. This strategy not only complements DR programs but also serves as a critical link between fluctuating market prices and consumer behavior.

The role of DR in mitigating market power is examined by \cite{devine2023role} using a market equilibrium model. They underscore the importance of distinguishing between regular consumers and prosumers with distributed energy resources (DERs). Such a distinction is crucial for a quantitative analysis of the effects and benefits of DR programs, illustrating the nuanced ways in which DR programs can influence market dynamics and consumer participation. 
For instance, \cite{jens2022classification} discuss approaches for leveraging prosumers' flexibility as balancing service providers in power systems, either through joint clustering or equilibrium settings. Prosumers can trade with other prosumers (\textit{peer-to-peer trading} (P2P)) or act as a larger unit within traditional electricity markets. One approach involves jointly clustering prosumers for a common goal within the microgrid, leveraging batteries, EVs, and heat pumps without explicitly modeling these devices. Another approach allows prosumers to act selfishly and exploit their flexibility in equilibrium, from which no participant can improve by unilaterally deviating from the equilibrium. 

 From a theoretical perspective, the demand response literature has made significant strides in addressing market complexities. It encompasses studies considering imperfect markets using bilevel models, such as \cite{afcsar2016achieving}, \cite{soares2019population}, \cite{aussel2020trilevel}, and \cite{soares2020designing} and perfect competition with equilibrium models, such as \cite{devine2023role,devine2018examining,hu2016stochastic,nguyen2016dynamic}.

 \cite{afcsar2016achieving} present a bilevel optimization model in which the supplier establishes time-differentiated electricity prices and consumers minimize their electricity consumption costs to enhance grid efficiency. \cite{soares2019population} present a model considering a single leader (retailer) whose aim is to maximize profit and multiple followers (consumers) whose objective is to minimize electricity costs. \cite{aussel2020trilevel} propose interactions among electricity suppliers, local agents, demand aggregators, and consumers with a tri-level multileader-multi follower game for load shifting induced by the time of use pricing.  \cite{soares2020designing} develop a comprehensive model in which a retailer maximizes its profit. A cluster of consumers responds to the retailer’s decisions on electricity prices by adjusting the operation of controllable loads to minimize their electricity bills and monetize a discomfort factor based on indoor temperature deviations. 
 
 Moreover, DR programs have been extensively analyzed using equilibrium models in various electricity market applications. The widespread application of these models in electricity market research stems from their robust ability to represent diverse market structures and interactions between market participants.
 For instance, \cite{devine2023role} investigate the role of DR in mitigating market power, particularly through load shifting and self-generation. \cite{devine2018examining} delve into the costs associated with providing flexibility through load shedding and self-generation using DERs. \cite{nguyen2016dynamic} analyze the energy scheduling challenges faced by load-serving entities in managing flexible and inflexible loads. \cite{hu2016stochastic} present a stochastic-multi-objective Nash-Cournot model to reduce peak demand, energy prices, and emissions.

 Furthermore, the application of tariff-based DR strategies has been explored to address other various challenges. These include distribution networks \citep{lu2018dynamic}, optimal bidding strategies in day-ahead electricity markets \citep{vaya2014optimal}, optimal pricing strategies in pool-based electricity markets \citep{ruiz2009pool}, modeling coordinated cyber-physical attacks \citep{li2015bilevel}, and optimal charging schedules for plug-in electric vehicles \citep{momber2015retail}, among others. Summarizing the benefits of DR programs from the perspective of various electricity market participants is a complex task, given the extensive range of literature on the subject. Some of the benefits are summarized in \cref{DR_benefits}.

\begin{figure}[H]
\centering
\begin{tikzpicture}[titlebox/.style={
        draw, thick, fill=blue!20, text width=.39\linewidth, align=center,
        minimum height=0.7cm, inner sep=2mm, font=\bfseries, anchor=north},
    contentbox/.style={draw, thick, fill=blue!5, text width=.39\linewidth, align=left,inner sep=5mm, anchor=north},
    overallborder/.style={draw, thick, inner sep=5mm, rectangle, rounded corners}]
\node[titlebox] (consumersTitle) {Consumers};
\node[titlebox, right=10mm of consumersTitle] (reliabilityTitle) {System Reliability};

\node[contentbox] (consumersContent) [below=0mm of consumersTitle.south] {
    - Incentive payments \\
    - System optimization \\
    - Improved comfort \\
    - Bill savings};
\node[contentbox] (reliabilityContent) [below=0mm of reliabilityTitle.south] {
    - Grid stability \\
    - Reduced volatility \\
    - Customer participation \\
    - Diversified resources};

\node[titlebox] (marketTitle) [below=5mm of consumersContent.south] {Electricity Market};
\node[titlebox] (environmentTitle) [below=5mm of reliabilityContent.south] {Environment};

\node[contentbox] (marketContent) [below=0mm of marketTitle.south] {
    - Deferred infrastructure costs \\
    - Price reduction \\
    - Capacity increase \\
    - Reduced market power};
\node[contentbox] (environmentContent) [below=0mm of environmentTitle.south] {
    - High renewable penetration \\
    - Rational use of energy resources \\
    - High economic efficiency \\
    - Carbon footprint reduction};
\node[overallborder, fit=(consumersTitle) (environmentContent)] {};
\end{tikzpicture}
\caption{Enhanced summary of DR program benefits.}
\label{DR_benefits}
\end{figure}
The literature on dynamic electricity price tariffs extensively examines their impact on individual players in the energy market. This includes cost minimization for demand aggregators, optimal reconfiguration of microgrids by DSOs, optimal allocation of distributed generations by TSOs, retailer profit maximization, and consumer benefit enhancement \citep{ajoulabadi2020flexible, wang2020smart, nejad2019reliability, dagoumas2017integrated, nilsson2018household}. 

By leveraging the advantages of game-theoretical models to address \textit{market power} and \textit{competitive markets}, in this work, we develop a tariff-based demand response program by simultaneously solving retailers' and consumers' problems. Assuming that consumers are rational decision makers and risk averse to their electricity consumption, we explicitly model their behavior by considering utility functions to quantify their welfare. 
Explicitly modeling human behavior is challenging, particularly in practical implementations, let alone determining its functional form. However, using utility functions to measure satisfaction is common in the classical literature, such as in performance evaluations of scheduling algorithms for telecommunication systems, where utility functions are widely used \citep{ palomar2007alternative}. 
The specific choice of a utility function to measure such behavior usually depends on the nature of the problem, the pattern of benefits under consideration, and the required mathematical properties, among other factors.
In demand response programs, utility functions can \textit{i)} be used as the statistical expectation of customers’ evaluation of electricity consumption \citep{samadi2012advanced}; \textit{ii)} illustrate how a rational consumer would make consumption decisions considering incentives from reducing/shifting electricity; and  \textit{iii)}  in practical settings, be inferred/imposed by tariffs under different market structures, such as those involving myopic and strategic retailers \citep{jens2022classification,niromandfam2020modeling}. Specifically, in settings where we can assume that consumers do not frequently switch to a different retailer (short-term model), the utility function should capture a decrease in its marginal value with the consumption level (\textit{diminishing returns}).  This could translate into demand curves where consumption levels decrease \textit{monotonically} with rising prices, as expected for rational consumers. Such demand behavior would also prevent retailers from excessively increasing price tariffs to maximize profit \citep{zugno2013bilevel}. Thus, the main contributions of this paper are:
\begin{enumerate}
   \item[\textit{i})] Applying demand response to address retailers' and consumers' decision-making challenges in electricity markets with dynamic pricing. Imperfect competition is analyzed using a Stackelberg game, and perfect competition is analyzed through equilibrium models, accounting for uncertainties by employing stochastic programming. 
\item [\textit{ii})] Tackling the bilevel problem between the retailer and consumers by directly solving the MPEC formulation without linearization, leveraging NLP solvers to handle the complexity of complementarity constraints. In our equilibrium model, consumer demand is treated as a variable in both upper- and lower-level problems, and dynamic prices are considered as exogenous parameters. The equilibrium value will naturally come out as the market clearing price that satisfies all the players’ (retailers’ and consumers’) optimality conditions. This approach contrasts with typical bilevel models, where price tariffs are treated as endogenous when transforming to MILP.
 \item [\textit{iii})] Introducing a utility function to quantify consumer flexibility, which, together with the availability of battery storage and other smart devices, extends consumer flexibility to two possibilities: \textit{i}) reacting to the price tariff where they can decide to buy more or less energy at a particular hour, and \textit{ii}) delaying or anticipating consumption within hours. 
 \item [\textit{iv})] Utilizing an extensive set of real-world case studies to illustrate the effectiveness of our model in capturing market structures and the implications of DR programs for the operational and economic decisions of market participants. The numerical results focus on analyzing the impact of consumer flexibility on social welfare and other market outcomes.
\end{enumerate}

 \section{Problem  formulation}\label{se_P_3_1}
 Most electricity markets exhibit a hierarchical decision-making structure that allows participants with market power to potentially anticipate the reactions of subsequent decision-making entities. In this regard, bilevel programming is an adequate optimization-based modeling technique for representing such interactions. It can be defined in various forms, including single-leader single follower, single leader multi follower, multi leader single-follower, and multi leader multi follower games. While the single leader- single follower case has been extensively analyzed since von Stackelberg's seminal work \citep{von1952theory}, its extensions, accounting for multiple followers, remain less explored.
 This work falls into the category where we model the problem faced by a single retailer with market power acting as a leader and multiple electricity consumers as followers. This scenario is realistic from the perspective of a single retailer aiming to set adequate price tariffs for consumers. Similarly, since consumers are \textit{attached} to a single retailer for their short-term decisions, we neglect potential interactions with rival retailers in capturing \textit{switching} consumers.
There are two common assumptions used to approach bilevel problems in the literature. \textit{i)} The \textit{optimistic approach} assumes that if a follower has more than one optimal solution, it will choose the solution that maximizes the leader’s objective. \textit{ii)} For any fixed decision of the retailer on the tariff, each consumer optimizes their behavior independently, without considering the challenges faced by rival consumers \citep{kovacs2019bilevel}. 
 
\begin{table}[t]
\caption{Main notations used in the model formulation for quick reference.}
\resizebox{\textwidth}{!}{%
\begin{tabular}{@{}ll@{}}
\toprule
\textbf{Notations} & \textbf{Description} \\ \midrule
\textbf{Sets} &  \\
$J$ & set of consumers ranging from $j=1,...,\mid J\mid$ \\
$ T$ & hours considered ranging from $t=1,...,|T|$ \\
$\Omega$ & set of scenarios ranging from $\omega =1,...,\mid\Omega\mid$ \\
\textbf{Parameters} &  \\
$A_{jt\omega}$ & linear coefficient of utility function for consumer $j$, hour $t$, scenario $\omega$ (\euro/kWh) \\
$B_{jt\omega}$ & quadratic coefficient of utility function for consumer $j$, hour $t$, scenario $\omega$ (\euro/kWh$^2$) \\
$C$ & penalization cost for power imbalances (\euro/kWh) \\
$\Delta_{j}^{C_{max}}$ & level of flexibility of consumer $j$ (kWh) \\
\textbf{Variables} &  \\
$\Delta_{jt\omega}^C$ & variation of consumer's demand $j$, hour $t$, scenario $\omega$ (kWh) \\
$q_{jt\omega}$ & electricity purchased by consumer $j$, hour $t$, scenario $\omega$ (kWh) \\
$q_{t\omega}^S$ & electricity purchased by retailer in the spot market, hour $t$, scenario $\omega$ (kWh) \\
$\delta_{t\omega}$ & electricity imbalance by retailer at hour $t$, scenario $\omega$ [kWh] \\
$P_t$ & retailer price tariff at hour $t$ (\euro/kWh) \\
$y_{t\omega}$ & electricity imbalance by retailer at hour $t$, scenario $\omega$ [kWh] \\
$P_{t\omega}^S$ & spot price at hour $t$, scenario $\omega$ (\euro/kWh) \\
\textbf{Dual Variables} &  \\
$\alpha_{t\omega}^+/\alpha_{t\omega}^-$ & upper/lower limits for electricity imbalance for retailer hour $t$, scenario $\omega$ \\
$\beta_{t\omega}$ & nonnegative electricity imbalance at hour $t$, scenario $\omega$ \\
$\lambda_{j\omega}$ & total quantity variation of demand for consumer $j$ under scenario $\omega$ \\
$\varepsilon_{jt\omega}$ & nonnegativity flexible and inflexible consumption of consumer $j$ at hour $t$, at scenario $\omega$ \\
$\mu_{t\omega}$ & electricity imbalance hour $t$, scenario $\omega$ \\
$\nu_{jt\omega}^{C_{max}}/\nu_{jt\omega}^{C_{min}}$ & maximum/minimum flexibility for consumer $j$ at hour $t$, scenario $\omega$ \\
$\theta_{t\omega}$ & nonnegative spot market electricity at hour $t$, scenario $\omega$ \\
$\gamma_{t} $ & nonnegative tariff decided by retailer at hour $t$ \\ \bottomrule
\end{tabular}%
}
\end{table}

Thus, we assume that a single retailer wants to optimize its expected profit, and $\left|J\right|$ consumers want to minimize the difference between the cost of purchasing electricity and the utility of their consumption. That is, the retailer determines the selling prices to maximize its expected profit, which depends on the consumers' purchases, and the consumers determine their load demands, which depend on the retail price tariffs. Consumers want to minimize disutility costs. First, we develop a bilevel programming problem assuming that retailers behave as leaders (upper-level) and that consumers behave as followers (lower-level). Then, we analyze the competitive equilibrium between the retailer and the consumers, assuming no market power. 
 
Bilevel programming with a strategic retailer is reformulated as an MPEC problem. This is done by recasting the consumers' problems using their KKT optimality conditions, which are introduced as constraints in the retailer's problem. Uncertainties related to consumer demand, spot market quantities and prices are addressed using stochastic programming.

For the equilibrium model with perfect competition, we formulate an MILP problem by linearizing the complementarity conditions of the corresponding retailer and consumers' KKT optimality conditions. In addition, an NLP reformulation of the equilibrium problem is derived using the methods proposed by \cite{leyffer2010solving}, which provides a solution equivalent to the MILP formulation.

The model has first-stage and second-stage decisions. Prior to day-ahead market clearance, the retailer decides the price tariff for hour $t$ (first stage). However, the spot market price, the quantity purchased by the retailer from the electricity market, the quantity bought by consumers, and the electricity quantity imbalance decisions are second-stage and scenario-dependent variables. 
As indicated, we seek to model rational behavior for consumers, via a utility function whose marginal value captures a monotonic decrease in consumption with price, as is observed in many markets, including electricity. For its analytical \textit{simplicity}, \textit{concavity}, and \textit{interpretability}, we chose a quadratic utility function of the form  $U_{jt\omega}(q_{jt\omega} + \Delta_{jt\omega}^C)= A_{jt\omega}(q_{jt\omega} + \Delta_{jt\omega}^C)-$ $\frac{1}{2}B_{jt\omega}(q_{jt\omega} + \Delta_{jt\omega}^C)^2$.  \textit{i}) It is simple to demonstrate \textit{market efficiency}, which helps understand how well the \textit{allocation rule} is designed to maximize the total utility of consuming $(q_{jt\omega} + \Delta_{jt\omega}^C)$ at hour $t$ by consumer $j$ under scenario $\omega$. This is achieved by solving the optimal amount of electricity, where the marginal utility of consuming an additional unit is zero, i.e, $q_{jt\omega} + \Delta_{jt\omega}^C =\frac{A_{jt\omega}}{B_{jt\omega}}$. To this end, the \textit{efficient allocation} of flexible and nonflexible electricity consumption for consumer $j$ at hour $t$ under scenario $\omega$ can be determined as $ \Delta_{jt\omega}^C =\frac{A_{jt\omega}}{B_{jt\omega}}- q_{jt\omega}$ and $q_{jt\omega} = \frac{A_{jt\omega}}{B_{jt\omega}}-\Delta_{jt\omega}^C$, respectively. \textit{ii}) Since $B_{jt\omega} > 0$, $U^{''}_{jt\omega}(q_{jt\omega} + \Delta_{jt\omega}^C)= -B_{jt\omega}$ is always negative, implying that the utility function is concave and demonstrates \textit{diminishing returns}. This means that each additional unit adds less utility to consumer $j$ at hour $t$ under scenario $\omega$ than the previous unit. 
Furthermore, to demonstrate that the overall implications of our proposed models are not solely dependent on our modeling choice, we also compare the results using a linear utility function in our analysis.

  \subsection{Retailer's problem}
The retailer maximizes its expected profit subject to technical and economic constraints, and accounts for uncertainty, which is modeled by a discrete number of scenarios. The retailer's problem is stated mathematically as follows:
\begin{subequations} \label{D_1}
        \begin{align}
        \displaystyle \max_{\bf{X}} & \sum_{\omega\in {\Omega}}\sigma_{\omega}\left(\sum_{j\in {J},t\in {T}}P_tq_{jt\omega}-\sum_{t\in {T}}(P_{t\omega}^Sq_{t\omega}^S+Cy_{t\omega})\right)\\
        &\textrm{subject to}\nonumber\\
        &\sum_{j\in {J}}q_{jt\omega}-q_{t\omega}^S = \delta_{t\omega}\quad:\mu_{t\omega}\quad \forall t,\forall \omega\label{1.1}\\
        &\delta_{t\omega}\le y_{t\omega}\quad: \alpha_{t\omega}^+\quad\forall t,\forall \omega\label{1.1.1}\\
        &-\delta_{t\omega}\le y_{t\omega}\quad:\alpha_{t\omega}^-\quad\forall t,\forall \omega\label{1.1.2}\\
         & y_{t\omega}\ge 0\quad: \beta_{t\omega}\quad\forall t,\forall \omega\label{1.1.3}\\
          & q_{t\omega}^S\ge 0 \quad:\theta_{t\omega}\quad\forall t,\forall \omega\label{1.2}\\
        & P_t\ge 0 \quad:\gamma_t \quad \forall t\label{1.3}
    \end{align}
    \end{subequations}
    where $\bf{X}$=$\{P_t, q_{t\omega}^S,\delta_{t\omega},y_{t\omega}\}$ is the retailer's set of decision variables, while $\mu_{t\omega},\alpha_{t\omega}^+,\alpha_{t\omega}^-$, $\theta_{t\omega},\gamma_t,\beta_{t\omega}$ are the dual variables associated with constraints (\ref{1.1})-(\ref{1.3}).\\
    The first term in the objective function represents the retailer's revenue from the quantity bought by consumer $j$ at hour $t$ under scenario $\omega$. This quantity is multiplied by the price $P_t$, which is the retail price of electricity (tariff) and a first-stage decision set by the retailer for each hour $t$. The second term is the cost of purchasing the electricity quantity $q_{t\omega}^S$ at the spot market price $P_{t\omega}^S$ at hour $t$ and with scenario $\omega$. The last term represents a cost, penalizing electricity imbalances. This cost can be positive or negative, where the variable $y_{t\omega}$ denotes the absolute value of these imbalances. The parameter $\sigma_{\omega}$ represents the probability associated with scenario $\omega$. The summation of the product over index $\omega\in \Omega$ represents the expected profit, which is what the retailer wants to maximize over the time horizon.
 
    Constraint (\ref{1.1}) is the electricity balancing constraint that guarantees the total demand by all consumers at hour $t$ with scenario $\omega$ minus the retailer's supply, which is the imbalance at that particular time. The linear sets of constraints (\ref{1.1.1}), (\ref{1.1.2}) and (\ref{1.1.3}) ensure that $y_{t\omega}=|\delta_{t\omega}|$. These imbalances are penalized in the objective function with weight $C$. Constraints (\ref{1.2}) and (\ref{1.3}) are nonnegativity constraints on the spot market quantity and price decided by the retailer for tariffs at hour $t$, respectively. $P_t$  is a first-stage decision that needs to be settled before uncertainty realization, while $q_{jt\omega}, q_{t\omega}^S, \delta_{t\omega}$, $P_{t\omega}^S$ and $y_{t\omega}$ are second stage decisions. The dual variables for each group of constraints are indicated on the right side of the colons. 
 \subsection{Consumers' problem}
In modeling consumer problems with DR, it is common to explicitly specify the type of consumer (residential, industrial, or commercial). This allows part of their consumption to be considered flexible, while another portion is inelastic and must be met at all times. 
 
 For instance, \cite{halvgaard2012economic} model heat pumps for heating residential buildings, where the heating system of the house becomes flexible consumption in the smart grid. \cite{zugno2013bilevel} extend this model by considering the heating dynamics of a building, the indoor temperature, or the temperature inside a water tank where flexibility (the output of interest) is characterized by the indoor temperature. They also consider the inflexible part of consumption in their model. However, in our model, including individual appliances would require a large number of additional variables and constraints. This would further complicate the numerical resolution and the economic interpretation of the results we aim to deliver. Moreover, we believe that the main demand patterns/profiles can be accounted for by the appropriate intertemporal adjustment of the utility function parameters. 
 
By considering the consumers' utility function explicitly in the objective function, we explore a residential consumer model with flexibility towards the dynamic price reported by the retailer. Thus, consumers face an economic problem where balancing the trade-off between electricity procurement costs and consumption utility is crucial. They must weigh the discomfort of deviating from their preferred consumption against the benefits of lower prices in a flexible pricing scheme. In particular, the consumers' problem can be represented within a game-theoretical setting, as it is parametrized by the decision of the retailer (hourly price $P_t$). The problem for consumer $j$ with scenario $\omega$ can be expressed as:
   \begin{subequations}\label{DR_2}
        \begin{align}
        \displaystyle \min_{q_{jt\omega},\Delta_{jt\omega}^C} & \sum_{t\in {T}}\left(P_tq_{jt\omega}-A_{jt\omega}(q_{jt\omega}+\Delta_{jt\omega}^C)+\frac{1}{2}B_{jt\omega}(q_{jt\omega}+\Delta_{jt\omega}^C)^2\right)\\
       & \textrm{subject to}\nonumber\\ 
       &q_{jt\omega}+\Delta_{jt\omega}^C\ge 0\quad:\varepsilon_{jt\omega}\quad \forall j,\forall t,\forall \omega \\
        & -\Delta_j^{C_{max}}\le \Delta_{jt\omega}^{C}\le +\Delta_j^{C_{max}}\quad:\nu_{jt\omega}^{C_{min}},\nu_{jt\omega}^{C_{max}} \label{2.1}\quad \forall j,\forall t,\forall \omega\\
        & \sum_{t\in {T}}\Delta_{jt\omega}^{C}=0\quad :\lambda_{j\omega}\quad \forall j,\forall \omega\label{d_4}
        \end{align}
    \end{subequations}
    where the objective function is the minus social welfare (cost minus utility) and $q_{jt\omega}$ (inelastic demand) and $\Delta_{jt\omega}^{C}$ (flexible demand) are consumer's decision variables. In the objective function $P_tq_{jt\omega}$ is the cost of purchasing $q_{jt\omega}$ electricity and $A_{jt\omega}(q_{jt\omega}+\Delta_{jt\omega}^C)-\frac{1}{2}B_{jt\omega}(q_{jt\omega}+\Delta_{jt\omega}^C)^2$ is the quadratic utility function that measures the benefit that consumer $j$ achieves by consuming the amount of energy ($q_{jt\omega}+\Delta_{jt\omega}^C$) during hour $t$, scenario $\omega$.
   
   Considering the consumers' objective function without any further constraints, consumption would only occur in periods when the real-time price is lower than the marginal benefit. Note that the marginal utility is a demand function $P_t=A_{jt\omega}-B_{jt\omega}(q_{jt\omega}+\Delta_{jt\omega}^C)$ where the electricity quantity demanded decreases as the electricity price increases. In other words, at the optimum, the marginal utility (benefit) from consuming electricity equals the retail price tariff $P_t$.

   Parameters $A_{jt\omega}$ and $B_{jt\omega}$ are the intercept and slope of the demand function, which are important parameters for characterizing the behavior of consumers toward the market outcomes of the model. Different values for these parameters can capture the dynamics of consumer demand. Constraint (\ref{2.1}) sets the lower and upper limits for flexible consumption, where $\Delta_j^{C_{max}}$ is the level of flexibility of consumer $j$. The inequalities guarantee that the flexible part of consumption at hour $t$ falls in the range between $-\Delta_j^{C_{max}}$ and $+\Delta_j^{C_{max}}$, which bound consumption increase and decrease, respectively. If $\Delta_j^{C_{max}}=0$ indicates that there is no flexibility in demand, while $\Delta_j^{C_{max}}>0$ indicates flexible demand. In the latter case, the consumer considers a dynamic price tariff that ensures cost savings and better welfare. In addition, constraint (\ref{d_4}) ensures that the net total flexible demand during the planning time horizon is zero. In the consumers' problem, $q_{jt\omega}$ and $\Delta_{jt\omega}^{C}$ are second-stage decisions. Note that since the dynamic electricity price $P_t$ enters the consumers' problem as a parameter (it is only a variable in the retailer problem), the optimization problems of consumers are convex (quadratic objective function with linear constraints). Therefore, we can replace each consumer's problem with its corresponding KKT optimality conditions \citep{haghighat2012bilevel}, which are sufficient for optimality.

\subsubsection{KKT formulation of the consumer problem}
The KKT optimality conditions of problem (\ref{DR_2}) are:
\begin{subequations}\label{6}
        \begin{align}
        &\frac{\partial\mathcal{L}_{j\omega}}{\partial q_{jt\omega}}= P_t-A_{jt\omega}+B_{jt\omega}(q_{jt\omega}+\Delta_{jt\omega}^C)-\varepsilon_{jt\omega}=0 \quad\forall j,\forall t, \forall\omega \label{second_1}\\
        &\frac{\partial\mathcal{L}_{j\omega}}{\partial \Delta_{jt\omega}^{C}}= -A_{jt\omega}+B_{jt\omega}(q_{jt\omega}+\Delta_{jt\omega}^C)- \nu_{jt\omega}^{C_{min}}+\nu_{jt\omega}^{C_{max}}
        -\lambda_{j\omega}-\varepsilon_{jt\omega}=0\quad\forall j,\forall t, \forall\omega \label{second_1.1}\\
         & \sum_{t\in {T}}\Delta_{jt\omega}^{C}= 0\quad\forall j,\forall \omega\label{second_5}\\
        &0\le q_{jt\omega}+\Delta_{jt\omega}^C\perp\varepsilon_{jt\omega}\ge 0\quad\forall j,\forall t,\forall \omega\label{cs_1}\\
        &0\le \Delta_{jt\omega}^{C}+\Delta_{j}^{C_{max}}\perp \nu_{jt\omega}^{C_{min}} \ge 0\quad\forall j,\forall t, \forall\omega\label{second_3}\\
         &0\le +\Delta_{j}^{C_{max}}-\Delta_{jt\omega}^{C}\perp\nu_{jt\omega}^{C_{max}}\ge 0\quad\forall j,\forall t,\forall \omega\label{second_4}\\
           &\nu_{jt\omega}^{C_{min}},\nu_{jt\omega}^{C_{max}},\varepsilon_{jt\omega}\ge 0\quad\forall j,\forall t, \forall\omega\label{second_10}\\
          &\lambda_{j\omega}\quad \text{free}\quad\forall j,\forall \omega\label{second_6}
        \end{align}
    \end{subequations}
where $\mathcal{L}_{j\omega}$ is the Lagrangian function for consumer $j$ and scenario $\omega$ of problem (\ref{DR_2}).
 By convention, the symbol $\perp$ indicates complementarity so that any of the two inequalities are satisfied as equality, i.e., the product of each expression and the corresponding dual variable must be zero.
 Note that the system of KKT optimality conditions is linear, with the exception of the complementarity conditions (\ref{cs_1})-(\ref{second_4}).
\subsection{ The Retailer's MPEC problem}
\begin{figure}
 \centering
 \includegraphics[scale=.82]{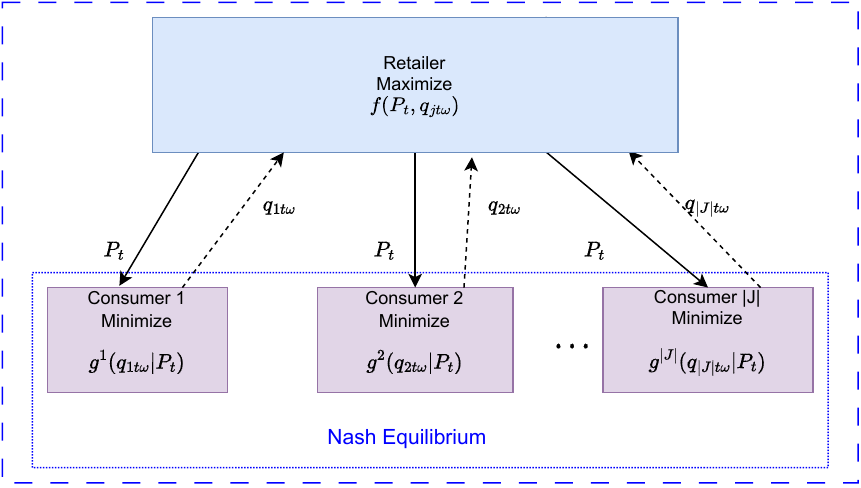}
 \caption{Game-theoretical framework for modeling demand response as a single leader-follower problem. The solid arrows indicate information signals from the retailer, while the dashed arrows represent inferences about consumers' behavior.}
 \label{Bilevel}
\end{figure}

In game theory, hierarchical optimization problems of this type can be formulated mathematically as MPECs. \cref{Bilevel} depicts the modeling framework for the MPEC problem where the leader optimizes his economic problem subject to his constraints, and by considering the equilibrium among the followers.

    Thus, the MPEC is expressed as follows:
     \begin{subequations}\label{mpec_last}
        \begin{align}
        \displaystyle \min_{\bf{X}^{MPEC}} & -\sum_{\omega\in {\Omega}}\sigma_{\omega}\left(\sum_{t\in{T},j\in {J}}P_tq_{jt\omega}-\sum_{t\in {T}}(P_{t\omega}^Sq_{t\omega}^S+Cy_{t\omega})\right)\label{M_1}\\
         &\textrm{subject to}\nonumber\\
         &(\ref{1.1})-(\ref{1.3})\\
          &(\ref{second_1})-(\ref{second_6})
        \end{align}
    \end{subequations}
    where $\bf{X}^{MPEC}$=$\{P_t,q_{t\omega}^S,\delta_{t\omega},y_{t\omega},q_{jt\omega},\Delta_{jt\omega}^c,\nu_{jt\omega}^{C_{max}},\nu_{jt\omega}^{C_{min}}, \varepsilon_{jt\omega},\lambda_{j\omega}\}$ is the set of retailer's decision variables.
    
    The MPEC is thus formulated as a single-level optimization problem, where the retailer minimizes his expected negative profit subject to other constraints and the KKT optimality conditions of the lower-level problems. However, the MPEC model in (\ref{mpec_last}) has two sources of nonlinearities: the bilinear product $P_tq_{jt\omega}$ in the objective function, and the complementarity conditions (\ref{cs_1})-(\ref{second_4}) in the consumers' KKT optimality conditions. 
    
The second type of nonlinearities can be linearized using the strategy proposed in \cite{fortuny1981representation}, which involves applying the big-M technique.  However, we still face the nonlinear product $P_tq_{jt\omega}$, which, as opposed to other problem settings \cite{ruiz2009pool} and \cite{sadat2017mixed}, cannot be linearized by employing the strong duality theorem. However, due to the observed good numerical performance, we have opted to solve the original MPEC problem with a state-of-the-art NLP solver such as KNITRO \citep{KNITRO}.

    \subsection{The equilibrium model} \label{sec:EquilibriumModel}
 In the case of perfect competition, we tackle the equilibrium model by jointly solving the KKT optimality conditions for retailers and consumers. This is achieved using two techniques: \textit{i}) by linearizing and casting the KKT systems of equations as a MILP problem, employing the Big-M technique from \cite{fortuny1981representation}, and \textit{ii}) by directly addressing and reformulating the problem as an NLP, following the approach in \cite{leyffer2010solving}. However, we mainly present the reformulation of the equilibrium problem as an MILP and its results, for our analysis. For details on the reformulation of NLP problems, the interested reader is referred to \cite{abate2021contracts}.
     \subsubsection{KKT formulation of the retailer problem}
To derive the retailer's KKT optimality conditions and address the equilibrium model, we assume a perfectly competitive market where players compete by choosing their optimal quantities. Consequently, $q_{jt\omega}$ also becomes a decision variable for the retailer. This setup implies that the price tariff $P_t$ is considered \textit{exogenous} for both retailers and consumers. Its equilibrium value will naturally emerge as the market-clearing price, i.e., the price that satisfies the optimality conditions of all players (retailers and consumers). 

The KKT optimality conditions derived from the retailer's problem are as follows:
\begin{subequations}\label{4}
        \begin{align}
        & \frac{\partial\mathcal{L}_{\omega}}{\partial q_{jt\omega}}= \sigma_{\omega}P_t+\mu_{t\omega}=0 \quad \forall j,\forall t,\forall \omega \label{second_b}\\
        & \frac{\partial\mathcal{L}_{\omega}}{\partial q_{t\omega}^S}= \mu_{t\omega}+\sigma_{\omega}P_{t\omega}^S-\theta_{t\omega}=0 \quad \forall t,\forall \omega \label{second_b.1}\\
    &\frac{\partial\mathcal{L}_{\omega}}{\partial \delta_{t\omega}}= \mu_{t\omega}+\alpha_{t\omega}^+-\alpha_{t\omega}^-=0\quad\forall t,\forall \omega \label{second_a}\\
     &\frac{\partial\mathcal{L}_{\omega}}{\partial y_{t\omega}}= \sigma_{\omega}C-\alpha_{t\omega}^+-\alpha_{t\omega}^--\beta_{t\omega}=0 \quad\forall t,\forall \omega \label{second_a.1}\\
     &\sum_{j\in {J}}q_{jt\omega}-q_{t\omega}^S = \delta_{t\omega}\quad\forall t,\forall \omega\label{new_1}\\
           &0\le y_{t\omega}-\delta_{t\omega} \perp\alpha_{t\omega}^+\ge 0\quad\forall t,\forall \omega\label{second_d.1}\\
           &0\le y_{t\omega}+\delta_{t\omega} \perp\alpha_{t\omega}^-\ge 0\quad\forall t,\forall \omega\label{second_d.2}\\
           &0\le q_{t\omega}^S \perp\theta_{t\omega}\ge 0\quad\forall t,\forall \omega\label{second_d}\\
         &0\le y_{t\omega} \perp\beta_{t\omega}\ge 0\quad\forall t,\forall \omega\label{second_e_c}\\
         & \alpha_{t\omega}^+,\alpha_{t\omega}^-,\beta_{t\omega},\theta_{t\omega}\ge 0\quad\forall t,\forall \omega\label{new}\\
         &\mu_{t\omega} \quad \text{free}\quad\forall t,\forall \omega.\label{second_e}
        \end{align}
    \end{subequations} 
 \subsection{Equilibrium problem formulation}
 Equilibrium problems in competitive markets can be approached in several ways. \textit{i)} As an MILP by linearizing complementarity constraints with the Big-M method. \textit{ii)} As an NLP, where products of the left- and right-hand sides of complementarity constraints form the objective function, with remaining feasibility and stationarity conditions as constraints. \textit{iii)} By linearizing complementarity constraints using a special order set type 1 (SOS type 1) method, which avoids arbitrary constants but increases computational cost. \textit{iv)} By formulating a mixed complementarity problem (MCP), which avoids Big-M values or nonlinear formulations by directly incorporating complementarity conditions and can be solved using tools like PATH or KNITRO. In this work, we choose the MILP approach with the Big-M method over MCP due to the broad availability and performance of MILP solvers, which align with practical industry applications. However, the Big-M method has limitations, as inappropriate values can introduce inaccuracies or computational issues \citep{pineda2019solving}. To address this, we select Big-M values based on capacity constraints and use a trial-and-error approach for Big-M values associated with bounds on dual variables. We also implement an NLP reformulation to validate our MILP results, which are consistent across both methods. 
We solve the equilibrium problem as MILP by concatenating the KKT optimality conditions ((\ref{second_1})-(\ref{second_6}) and (\ref{second_b})-(\ref{second_e})). The complementarity slackness conditions (\ref{cs_1})-(\ref{second_4}) and (\ref{second_d.1})-(\ref{second_e_c}) are linearized by introducing an auxiliary variable for each condition.

Then, to exploit the efficient performance of MILP solvers, the resulting system of MIL constraints can be reformulated as an MILP by including an arbitrary constant as  the objective function: 
 \begin{subequations}\label{mix}
        \begin{align}
        &\displaystyle \min_{\bf{X}^{Comp}} 1\\
        &\textrm{subject to}\nonumber\\
         &(\ref{second_1})-(\ref{cs_1})\\
         &(\ref{second_b})-(\ref{new_1}), (\ref{new})\\
         &(\ref{lin_4})-(\ref{lin_5})\\
         &(\ref{u_1})-(\ref{U_2})\\
         &(\ref{U_3})-(\ref{U_4}) \\
          &\mu_{t\omega}\quad\text{free}\quad \forall t,\forall \omega
           \end{align}
    \end{subequations}
    where $\bf{X}^{Comp}$=$\{q_{t\omega}^S,\delta_{t\omega},y_{t\omega}, q_{jt\omega},\Delta_{jt\omega}^C,\nu_{jt\omega}^{C_{min}},
    \nu_{jt\omega}^{C_{max}}, \varepsilon_{jt\omega},\lambda_{j\omega}, \mu_{t\omega},\alpha_{t\omega}^+,\alpha_{t\omega}^-,\theta_{t\omega},\beta_{t\omega}$,\\ $\psi_{jt\omega},\psi_{jt\omega}^{C_{min}},\psi_{jt\omega}^{C_{max}}, \tau_{t\omega}^+,\tau_{t\omega}^-, \tau_{t\omega}^q,\tau_{t\omega}^y,\tau_{t}^P\}$ is the set of decision variables for the competitive model.
In particular, the complementarity conditions from the consumer's KKT conditions can be linearized as follows:
        \begin{subequations}\label{li_2}
        \begin{align}
      &\Delta_{jt\omega}^C+\Delta_j^{C_{max}}\ge 0, \quad\forall j,\forall t,\forall \omega\label{lin_4}\\
      &\Delta_j^{C_{max}}-\Delta_{jt\omega}^C\ge 0, \quad\forall j,\forall t,\forall \omega\\
       &q_{jt\omega}+\Delta_{jt\omega}^C\ge 0\quad\forall j,\forall t,\forall \omega\\
      &\nu_{jt\omega}^{C_{min}}\ge 0\quad\forall j,\forall t,\forall \omega\\
       &\nu_{jt\omega}^{C_{max}}\ge 0\quad\forall j,\forall t,\forall \omega\\
       &\varepsilon_{jt\omega}\ge 0\quad\forall j,\forall t,\forall \omega\\
        &\Delta_{jt\omega}^C+\Delta_j^{C_{max}}\le (1-\psi_{jt\omega}^{C_{min}})M^c, \quad\forall j,\forall t,\forall \omega\\
         &\Delta_j^{C_{max}}-\Delta_{jt\omega}^C\le (1-\psi_{jt\omega}^{C_{max}})M^c, \quad\forall j,\forall t,\forall \omega\\
         &q_{jt\omega}+\Delta_{jt\omega}^C\le (1-\psi_{jt\omega})M^c\quad\forall j,\forall t,\forall \omega\\
         &\nu_{jt\omega}^{C_{min}}\le \psi_{jt\omega}^{C_{min}}M^{P}, \quad\forall j,\forall t,\forall \omega\\
         &\nu_{jt\omega}^{C_{max}}\le \psi_{jt\omega}^{C_{max}}M^{P}, \quad\forall j,\forall t,\forall\omega\\
         &\varepsilon_{jt\omega}\le \psi_{jt\omega}M^{P}, \quad\forall j,\forall t,\forall\omega\\
         &\psi_{jt\omega},\psi_{jt\omega}^{C_{min}},\psi_{jt\omega}^{C_{max}} \in \{0,1\}\quad\forall j,\forall t,\forall\omega\label{lin_5}
        \end{align}
    \end{subequations}
    where $M^c$ and $M^{P}$ are sufficiently large constants.
    
         (\ref{second_d.1}) and (\ref{second_d.2}) are linearized as follows:
         \begin{subequations}\label{mix_1}
        \begin{align}
         &y_{t\omega}-\delta_{t\omega}\ge 0,\quad \forall t, \forall\omega\label{u_1}\\
         &y_{t\omega}+\delta_{t\omega}\ge 0, \quad\forall t, \forall\omega\\
         &\alpha_{t\omega}^+\ge 0,\quad\forall t, \forall\omega\\
          &\alpha_{t\omega}^-\ge 0,\quad\forall t, \forall\omega\\
          &y_{t\omega}-\delta_{t\omega}\le(1-\tau_{t\omega}^+)M^S\quad\forall t, \forall\omega\\
          &y_{t\omega}+\delta_{t\omega}\le(1-\tau_{t\omega}^-)M^S\quad\forall t, \forall\omega\\
           &\alpha_{t\omega}^+\le\tau_{t\omega}^+M^{T},\quad\forall t, \forall\omega\\
           &\alpha_{t\omega}^-\le\tau_{t\omega}^-M^{T},\quad\forall t, \forall\omega\\
           &\tau_{t\omega}^+,\tau_{t\omega}^- \in \{0,1\}\label{U_2}
              \end{align}
    \end{subequations}
     where $M^S$ and $M^{T}$ are sufficiently large constants.
     
    Finally, the linearization of (\ref{second_d})-(\ref{second_e}) renders:
           \begin{subequations}\label{mix_2}
        \begin{align}
           &q_{t\omega}^S\ge0\quad\forall t,\forall\omega\label{U_3}\\
            &y_{t\omega}\ge0\quad\forall t,\forall\omega\\
            &\theta_{t\omega}\ge0\quad\forall t,\forall\omega\\
              &\beta_{t\omega}\ge0\quad\forall t,\forall\omega\\
                &q_{t\omega}^S \le(1-\tau_{t\omega}^q)M^S,\quad\forall t,\forall\omega\\
           &y_{t\omega}\le(1-\tau_{t\omega}^y)M^S,\quad\forall t,\forall\omega\\
               &\theta_{t\omega}\le\tau_{t\omega}^qM^{T},\quad\forall t,\forall\omega\\
              &\beta_{t\omega}\le\tau_{t\omega}^yM^{T},\quad\forall t,\forall\omega\\
               &\tau_{t\omega}^q,\tau_{t\omega}^y,\tau_{t}^P\in\{0,1\}\label{U_4}
        \end{align}
    \end{subequations}
    where $M^S$ and $M^{T}$ are sufficiently large constants.
           \section{Numerical results and discussion}\label{se_P_3_2}
  This section presents numerical simulations and discusses the models' practical implications. It also analyzes the models' performance under various market conditions and demand response scenarios, with sensitivity analyses to ensure the robustness of the results. The numerical settings for the simulations, including data sources, parameter values, and scenario generation are also discussed. These settings are based on standard practices in the literature and are validated against real-world data from the European Energy Exchange platform. Additionally, we perform sensitivity analyses on some key model parameters to confirm the robustness and reliability of our findings.

    \subsection{Data}
 We consider a day-ahead electricity market clearing price from the European Energy Exchange (EEX).  The day-ahead electricity market clearing prices are available in one-hour intervals and can be found at \cite{EEX2023}. 
 The marginal utility is the price-demand curve, where $A_{jt\omega}$ and $B_{jt\omega}$ are the price-demand curve intercept and slope, respectively. By utilizing day-ahead clearing price and quantity data from a specific date, we can calibrate the expected values for utility parameters ($A_{jt\omega}$ and $B_{jt\omega}$) for consumer $j$, at time $t$ under scenario $\omega$. Subsequently, introducing perturbations allows the levels of flexibility among residential consumers to vary \citep{zugno2013bilevel}. Regarding consumers’ flexibility, $\Delta_{j}^{C_{max}}= 0$ indicates \emph{no flexibility}, while $\Delta_{j}^{C_{max}} > 0$ signifies consumers' \emph{flexibility} to varying degrees. 

Although theoretically there is no limit to the number of consumers, we simplify our model by limiting it to three consumers.  This limitation to three consumers is sufficient to draw insights from the considered model, and \textit{i}) the findings can be generalized to representative real-world scenarios \citep{niromandfam2020modeling,zugno2013bilevel}.
\textit{ii}) We address the preferences of three consumers by adjusting demand parameters and offering flexibility in response to changes in consumption. This entails analyzing the cases of each consumer both before and after the implementation of the demand response program. This means that expected values of parameters $A_{j}$, $B_{j}$, and $\Delta_{j}^{C_{max}}$ are assumed to vary. This reflects the reactions of consumers to the retailer's price signals while competing with each other to impress the retailer and by maximizing their utility(minimizing their disutility). Consumer $1$ demonstrates the lowest slope and greatest flexibility, while Consumer $3$ shows the least flexibility, and Consumer $2$ lies between these two extremes (as detailed in \cref{Tab_1}). This helps us to understand the impact of the parameters on market outcomes and their dynamics.   
\begin{table}[t]
\centering
\caption{Mean values of consumer parameters in the six case studies ($j= 1,2,3$): $A_j$, $B_j$, and $\Delta_{j}^{C_{max}}$ are expressed in units of \euro/kWh, \euro/kWh$^{2}$, and kWh, respectively.}
\resizebox{\textwidth}{!}{%
\begin{tabular}{|l|l|l|l|l|l|l|l|l|l|}
\hline
& \multicolumn{9}{c|}{Parameters {}{}} \\ \cline{2-10}
Cases  & $A_1 $ & $A_2$ & $A_3$ & $B_1$ & $B_2$ & $B_3$ & $\Delta_{1}^{C_{max}}$ & $\Delta_{2}^{C_{max}}$ & $\Delta_{3}^{C_{max}}$ \\ \hline
Benchmark & 0.0291 & 0.0302 & 0.0271 & 0.0013 & 0.0015 & 0.0014 & 2.50 & 1.40 & 2.00 \\
Linear  & 0.035 &0.0375 &0.0341 &0.0013 &0.0015 &0.0014 &2.50 &1.40 &2.00 \\
Quadratic &0.0291 &0.0302 &0.0271 &0.0017 &0.020 &0.0019 &2.50 &1.40 &2.00 \\
Flexibility &0.0291 &0.0302 &0.0271 &0.0013 &0.0015 &0.0014 &5.00 &2.40 &3.50 \\ 
No Flexibility &0.0291 &0.0302 &0.0271 &0.0017 &0.020 &0.0019 &0.00 &0.00 &0.00 \\
Linear utility &0.0291 &0.0302 &0.0271 &0.00 &0.00 &0.00 &5.00 &2.40 &3.50 \\ \hline
\end{tabular}}
\label{Tab_1}
\end{table}
\subsection{Case studies}

 We consider six cases for both the MPEC and MILP (equilibrium) models in the numerical analysis:
\begin{itemize}
    \item [\textit{i})] \emph{Benchmark}: This case serves as a reference for the increase or decrease in demand parameters ($A_{jt\omega}, B_{jt\omega}, \Delta_{j}^{C_{max}}$) calibrated using the input data from \cref{Tab_1}.
\item [\textit{ii})] \emph{Linear}: Parameter $A_{j}$ is adjusted by a 25\% increase compared to its value in the \emph{benchmark} case while keeping the other parameters constant.
\item [\textit{iii})]\emph{Quadratic}: Parameter $B_{j}$ is adjusted by a 35\% increase compared to its value in the \emph{benchmark} while keeping the other parameters constant.
\item [\textit{iv})] \emph{Flexibility}: The parameter $\Delta_{j}^{C_{max}}$ is adjusted by an 82\% increase compared to its value in the \emph{benchmark} case while keeping the other parameters constant.
\item [\textit{v})] \emph{No flexibility}: The parameter $\Delta_{j}^{C_{max}}$ is set to $0$.
\item [\textit{vi})]\emph{Linear utility}: The parameter $B_{jt\omega}$ is set to $0$, and the other parameters are kept constant at the \emph{benchmark} values.

\end{itemize}
Based on the cases, we illustrate and compare the market outcomes across these case studies.

 To account for uncertainty, we implement stochastic programming with equiprobable scenarios and apply Gaussian distributions to the parameters. We then utilize the mean values in \cref{Tab_1} to calibrate the data and select the coefficients of variation (CVs) for the parameters to determine the corresponding standard deviations. For instance, to compute the standard deviation of spot prices, we use the formula $CV \times E[P_{t\omega}^S]$. Thus, $E[P_{t\omega}^S]$ represents the expected spot price at time $t$ under scenario $\omega$, and we set the CV at $0.015$ for the spot market calculation.  $E[P_{t\omega}^S]$ is the day-ahead clearing prices for $24$ hours under scenario $\omega$. $P_{t\omega}^S$—the spot price under scenario $\omega$—is used for the simulation to reflect $\omega$ possible scenarios regarding market clearing price dynamics. Similarly, we generate $A_{jt\omega} \sim N(\mu_{A_{jt\omega}},\sigma_{A_{jt\omega}})$ and $B_{jt\omega} \sim N(\mu_{B_{jt\omega}},\sigma_{B_{jt\omega}})$, where their standard deviations are calculated using CV values of 0.013 and 0.0013, respectively. Finally, the demand flexibility parameter $\Delta_{j}^{C_{max}}$ can vary from 0 to 5.0 kWh. This range is chosen because it reflects realistic levels of flexibility for individual consumers participating in DR programs.
  
The models were solved using JuMP version 1.11.1 \citep{Lubin2023} under the open-source Julia programming language version 1.8.5 \citep{bezanson2017julia}. We use Artelys KNITRO solver version 13.2 \citep{byrd2006k} for the MPEC and Gurobi version 10.1 \citep{gurobi} for the MILP on a CPU E5-1650v2@3.50GHz and 64.00 GB of RAM running workstation. For the MPEC problem, we test the case study with $\Omega= 30$ scenarios, and for the MILP problem $\Omega= 300$. However, in the MPEC model, the number of scenarios increases, and the number of variables and constraints in the MPEC problem increases exponentially. This renders the problem computationally intractable, especially when the number of scenarios exceeds 30($\Omega > 30$)\footnote{We do not present computational considerations by varying the number of scenarios for the sake of space.}. There are possible alternatives to tackle this computational challenge, such as utilizing high-performance computing (HPC) with parallel computing techniques (we refer the interested reader to \cite{ahmadi2013multi} and \cite{nasri2015network}). Since our problem is nonconvex as it involves solving the MPEC problem without applying any linearization or approximations, we tackled it by using the NLP solver KNITRO. Although the presented solutions are denoted as optimal by the solver, proving that their global optimality is not analytically possible, we addressed this issue numerically by employing a multistart strategy. This approach involves running the solver from multiple initial points, whereby the algorithm selects the best solutions after discarding potential lower-quality optima.

Moreover, a sensitivity analysis on the number of scenarios shows that increasing the number of scenarios has no significant impact on market outcomes (see \cref{Tab_3}). For the equilibrium problem with MILP, there is no computational issue concerning the number of scenarios. The model can be efficiently solved for scenarios, such as  $\Omega=500$ or greater. The MPEC requires $42.422 $ of CPU time with $\omega = 30$, while the MILP requires $15.019$ of CPU time with $\omega = 300$. It is also important to note that the methods to adjust big-Ms are mostly heuristic and problem-dependent, where the trial-and-error procedure has been used in most related research \citep{motto2005mixed,garces2009bilevel,jenabi2013bi}. However, for the particular problem addressed, the tuning of these constants is achieved without particular numerical trouble. The primary challenge lies in selecting appropriate Big-M values associated with the dual variables. These values must be large enough to avoid imposing unnecessary constraints on the problem. However, they must not be excessively large, as that would risk creating an ill-conditioned model \citep{pineda2019solving}.

\begin{table}[t]
\centering
\caption{Market outcome sensitivity to the number of scenarios in the MPEC model under the \emph{benchmark} case.}
\resizebox{\textwidth}{!}{%
\begin{tabular}{|l|l|l|l|l|l|l|l|l|l|}
\hline
   & \multicolumn{4}{c|}{Market outcomes} &    & \multicolumn{4}{c|}{Market outcomes} \\ \cline{2-5} \cline{7-10} 
$\Omega$ & Expected profit & $P_t$ & $q_{t\omega}^S$ & $q_{jt\omega}$ & $\Omega$ & Expected profit & $P_t$ & $q_{t\omega}^S$ & $q_{jt\omega}$ \\ \hline
10 & \multicolumn{1}{l|}{0.1857} & \multicolumn{1}{l|}{0.0268} & \multicolumn{1}{l|}{14.047} & 4.682 & 20 & 0.1839  & 0.0267   & 14.211  & 4.737 \\ 
15 & \multicolumn{1}{l|}{0.1850} & \multicolumn{1}{l|}{0.0268} & \multicolumn{1}{l|}{14.074} & 4.691 & 30 & 0.1844  & 0.02267  & 14.201  & 4.733 \\ \hline
\end{tabular}}
\label{Tab_3}
\end{table}
 
 \subsection{Market outcomes: Electricity price tariffs}
\begin{figure}
 \centering
 \subfloat[Comparison of price tariffs in both models across various cases and the day-ahead clearing prices.]{\includegraphics[scale=.84]{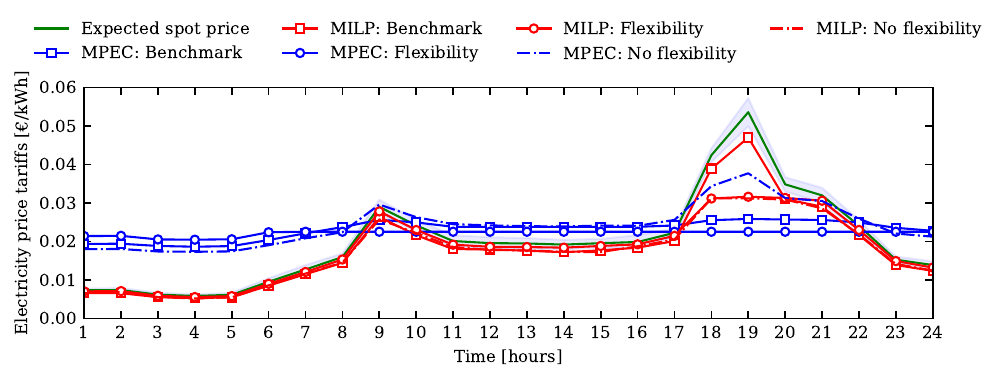}}\\
\subfloat[Variation in price tariffs with MPEC by adjusting model parameters.]{\includegraphics[scale=.840]{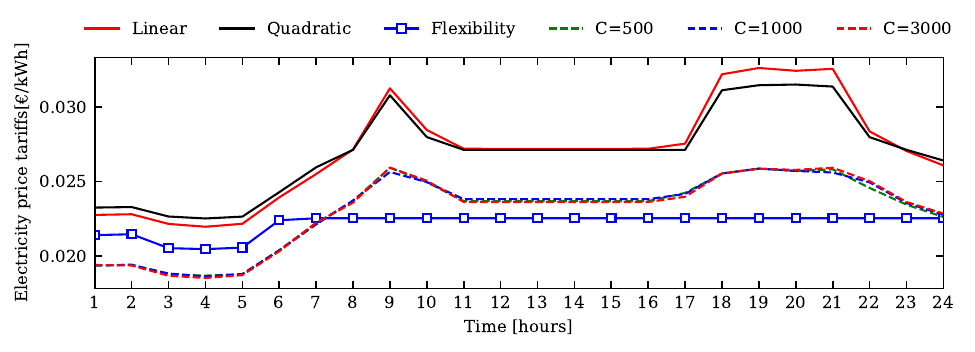} 
 \label{Prices}}
 \caption{Comparison of electricity price tariffs: MPEC vs. MILP across various cases.}
\end{figure}

 The retailer's price tariffs in both the MPEC and the MILP models are depicted along with the expected day-ahead electricity market prices in \cref{Prices}(a). 
The price tariffs are compared across the \emph{benchmark}, \emph{flexibility}, and \emph{no flexibility} \footnote{Note that we use \textit{No flexibility} to refer to the scenario \textit{before DR} implementation and \textit{flexibility} to describe the case \textit{after DR}} implementation. cases in both models. In all cases, the MILP price tariffs demonstrate consistent patterns with the market clearing prices in the day-ahead market.
However, with the MPEC model, when the retailer's market power is considered, price tariffs exceed the expected day-ahead electricity prices even during low-demand periods. This is market inefficiency, primarily due to market distortion caused by the strategy of the retailer.
 On the other hand, within the competitive equilibrium model (MILP), price tariffs are generally lower than those in the MPEC model across most cases and periods. The overall price tariffs $P_t$ set by the retailer under the MPEC framework are consistently higher than the spot market prices $P_{t\omega}^S$. Furthermore, the retailer reduces its procurement when the spot market prices are relatively high (at $9:00, 10:00, 17:00-21:00$). It increases its purchases when the spot market prices are lower (at $t=$1:00-8:00,10:00-17:00). These electricity price patterns align with the quantities of power the retailer purchases during peak periods, which is reflected in the optimal outcomes showing lower quantities in the spot market. A more detailed analysis is available in \cref{PP_2}. It provides a comprehensive comparison of results before and after implementing the demand response program under the MPEC and MILP models. Higher quantities are purchased when wholesale market prices are lower. Although lower electricity price tariffs are already in place during the early hours, demand flexibility further decreases electricity price tariffs in the MPEC model. In the equilibrium model (MILP), the overall electricity price tariffs remain below the spot price outside the peak periods.
    
Consumer flexibility results in lower retailer price tariffs in the MPEC, especially after 7:00. This means that if consumers adjust their consumption, they can further reduce their electricity prices. Otherwise, prices can increase with less flexibility. The MPEC model also has a peak-shaving and valley-filling effect. In cases of high flexibility, the realized price tariffs during on-peak and off-peak periods become very similar. This is a much-desired result, as a flatter price tariffs curve means that the system is more predictable and reliable. In the MILP model, however, high demand elasticity reduces electricity prices and increases the quantities purchased in both the spot and retail markets more than in the MPEC model. Thus, without demand flexibility, price tariffs are greater in the \emph{benchmark} under the MPEC model than in the MILP model. Compared with those in the  \emph{flexibility} case, prices do not significantly change over time in the \emph{benchmark} case. 

\cref{Prices}(b) illustrates the impact of increasing demand parameters on price tariffs. While an increase in $A_{jt\omega}$ slightly increases the clearing price, the overall impact of demand parameters and the penalty weight on price tariffs is insignificant. Note that imposing a penalty on the retailer's objective function affects the power imbalance in the periods considered rather than its impact on electricity price tariffs. Interestingly, increasing the flexibility parameter $\Delta_{j}^{C_{max}}$ leads to a significant reduction in price tariffs, more than other model parameters. A key takeaway is that market power hinders the efficiency of consumer flexibility, where the price-maker retailer exploits consumer flexibility to maximize its expected profits by signaling higher tariffs.

\subsection{Market outcomes: Loads and spot market qunatities}

   \begin{figure}
 \centering
  \subfloat[Comparison of purchased quantities in both models before and after the implementation of the demand response program.]{\includegraphics[scale=.84]{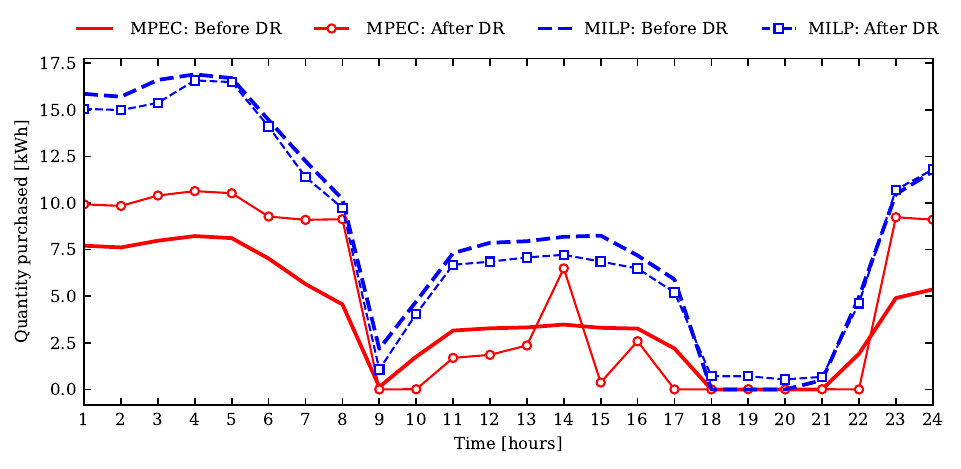}}\\
   \subfloat[Quantities purchased in spot market for both models before and after the implementation of the demand response program.]{\includegraphics[scale=.58]{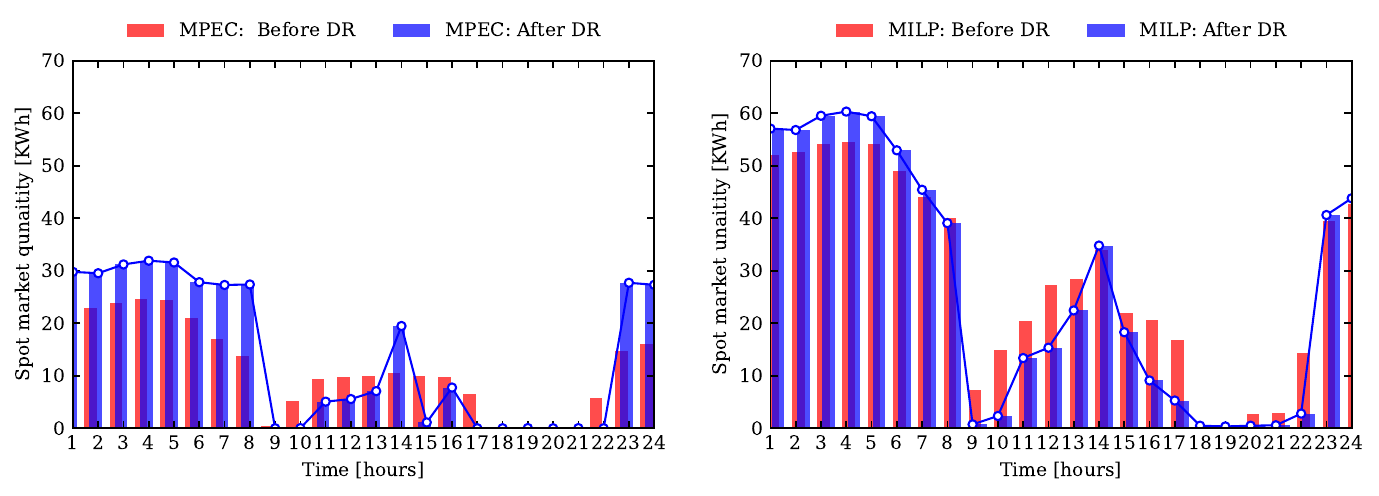}}\\
   \caption{Comparison of consumption and spot market quantities for the \emph{no flexibility} and \emph{flexibility} cases.}
 \label{PP_2}
\end{figure}
 
\cref{PP_2}(a) depicts the aggregate consumer purchases in the retail market both before and after the implementation of the demand response in the MPEC and MILP models. The results indicate that lower spot market prices lead to increased purchases by consumers in the MILP model. The trend is similar without the implementation of the demand response program.
With MILP, a higher value of $B$ (\emph{quadratic} case) is associated with significant and consistent electricity purchases in the retail market. Note that an increase in the quadratic term of the utility function is analogous to a decrease in electricity prices. Both parameters are inversely related to the quantity of electricity purchased. In contrast, an increase in the parameter in $A$ (\emph{linear} case) encourages consumers to buy more electricity during off-peak periods in both the MPEC and MILP models, thereby avoiding higher tariff periods. Consequently, lower prices during off-peak periods prompt higher purchases from the spot and retail markets. As loads shift from expensive to cheaper periods using consumer flexibility, new peak and valley periods emerge. This indicates that a peak period is typically followed by a valley period. Note that the detailed results for the \emph{linear} and \emph{quadratic} cases are omitted for brevity.

 With different cases, the retailer buys higher quantities in perfect competition where there is a lower price, particularly in the \emph{quadratic} case  (higher $B$ values). With the \emph{benchmark} case, where there are lower values for $A$ and $\Delta_{j}^{C_{max}}$, the retailer purchases a lower quantity even in the MILP model. The equilibrium model shows that the greater the competition is, the lower the impact the retailer could have on the market outcomes. Specifically, the price tariffs sent by the retailer are lower in the MILP model because the retailer is not a price-maker. This enables consumers to enjoy greater benefits when purchasing their electricity in a competitive market model.

\cref{PP_2}(b) shows the quantities purchased in the spot market with MPEC and MILP under the \emph{benchmark} case.
The spot market quantities are consistent in both models before and after the DR program was implemented. However, with high \emph{flexibility}, the quantities in the spot market during off-peak periods are significantly greater. This shows that consumer flexibility allows retailers to shift their purchases to off-peak periods to benefit from lower prices in the wholesale market. During peak demand periods, retailers might decide not to buy from the day-ahead market due to the high spot market prices. If retailers have to purchase electricity during peak periods when wholesale prices are high, they can send higher price tariffs to consumers during these periods. This further motivates consumers to shift/shave their consumption, which benefits both parties.

A comparison of consumers' purchased quantities in the \emph{benchmark} case shows individual consumer behaviors based on their utility function parameters. Consumer $1$ tends to buy more than other consumers when electricity price tariffs are low, such as in the early morning and late evening. This is because Consumer $1$ has the minimum slope  ($A_{1tw} = 0.0013$) and the highest demand flexibility ($\Delta_{j}^{C_{max}}=2.50 $) in the \emph{benchmark} case. Individual consumers' results are omitted for brevity.

\begin{figure}
 \centering
 \subfloat[Price-quantity comparison: \emph{benchmark}.]{\includegraphics[scale=.4]{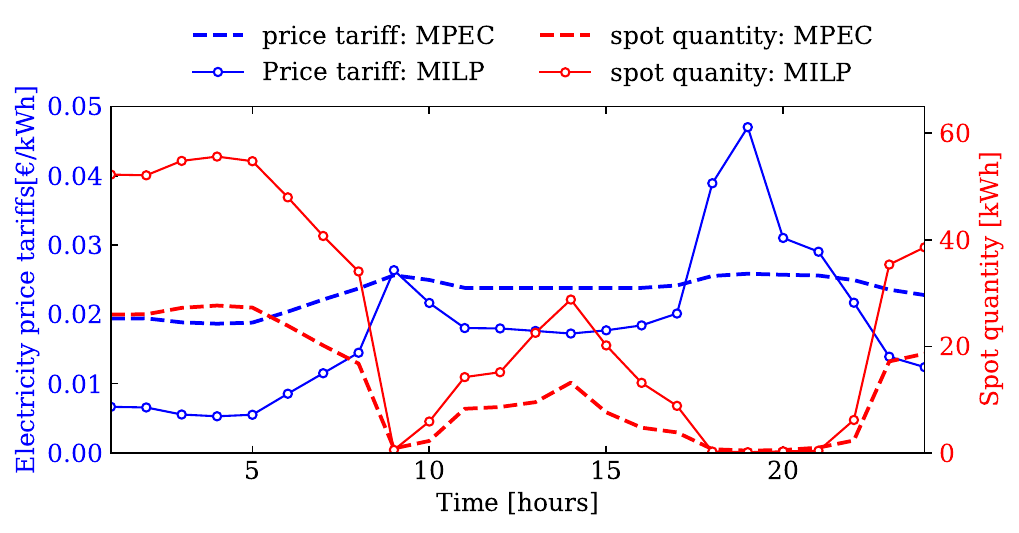}}
 \subfloat[Price-quantity comparison: \emph{linear}.]{\includegraphics[scale=.4]{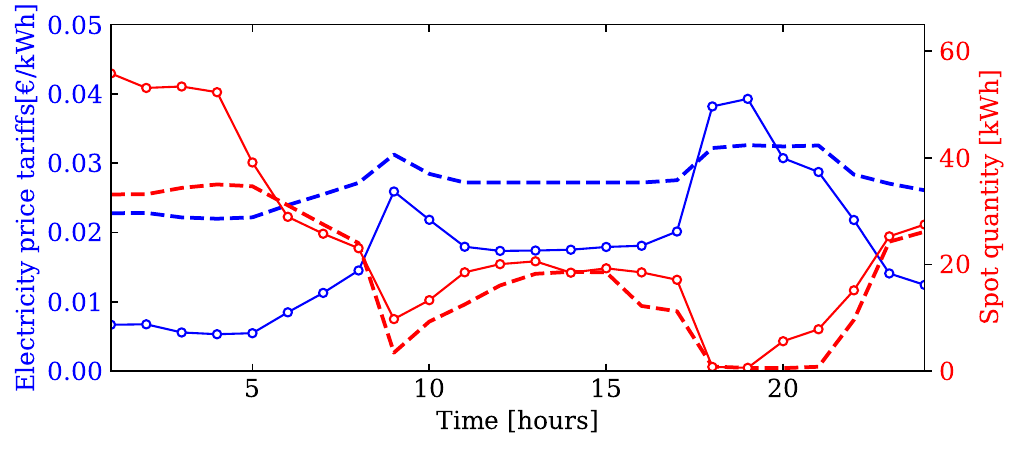}}\\
  \subfloat[Price-quantity comparison: \emph{quadratic}.]{\includegraphics[scale=.4]{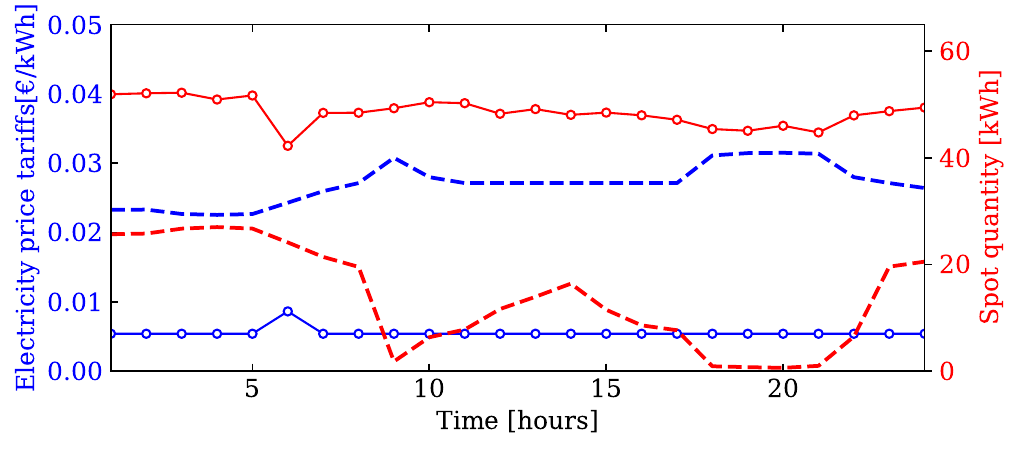}}
  \subfloat[Price-quantity comparison: \emph{flexibility}.]{\includegraphics[scale=.4]{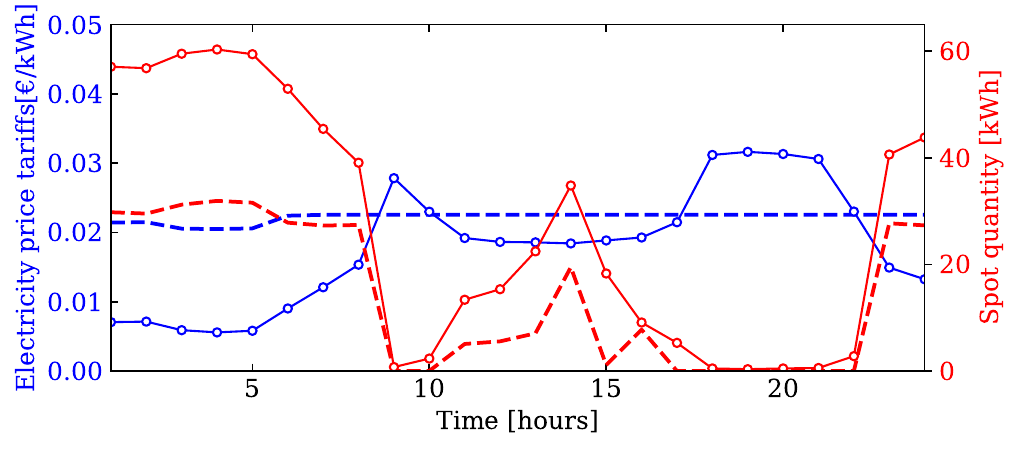}}\\
 \caption{Price (left axis) and spot quantity (right axis) comparison in both models and for various cases (\emph{benchmark, linear, quadratic} and \emph{flexibiltiy}).}%
 \label{P_3_7}%
\end{figure}
\subsection{Market outcomes: Electricity prices and retailer's quantities comparison}
Now, let us compare price tariffs, quantities, consumers' utility parameters, and demand flexibility to understand the dynamics of the market.
\cref{P_3_7} depicts the price tariffs sent by the retailer to consumers and quantities purchased by the retailer from the spot market in both models. Price tariffs are plotted on the left axis, and spot market quantities are plotted on the right axis. \cref{P_3_7}(a) depicts the \emph{benchmark}  case, which serves as a baseline for comparing the impacts of the market parameters on market outcomes. In the quadratic utility function, the linear term (as shown in \cref{P_3_7} (b)) captures the initial positive impact of electricity consumption on consumers' utility. Moreover, the \emph{quadratic} case models decrease in utility as electricity consumption increases. As a result, when the parameter $B_{jt\omega}$ increases, the retailer purchases a larger quantity from the wholesale market. This increase is illustrated in \cref{P_3_7}(c). The retailer then sells this quantity to consumers at lower price tariffs in the retail market. A lower price, in turn, increases consumer utility. It should be noted that from the retailer's perspective, an increase in $B_{jt\omega}$ has an effect analogous to an increase in demand from consumers.
 
Overall, the electricity price tariffs sent by the retailer and the quantities bought by the retailer from the spot market exhibit similar trends. In a competitive market, consumers' utility increases due to lower prices. In the MPEC market, the expected profit of the retailer increases as it strategically affects price formation. While prices remain relatively stable in the MPEC model, the quantities purchased from the spot market are significantly affected by the flexibility of consumers. Most notably, greater flexibility has a greater impact on the quantity purchased in the spot market. A key takeaway is that as consumers become more willing to shift their consumption to periods of lower demand, the retailer has to buy from the wholesale market to fulfill demands during off-demand periods. This shift leads to an overall decrease in wholesale electricity prices and enhances overall social welfare as detailed in \cref{CW}. From these simple simulation results, the managerial insight is that market regulators should consider the market structure when designing an efficient demand response program to exploit the potential of consumer flexibility for both spot and retail market outcomes.

\subsubsection{Market outcomes: expected profit}
\begin{figure}
\centering
\includegraphics[scale=.850]{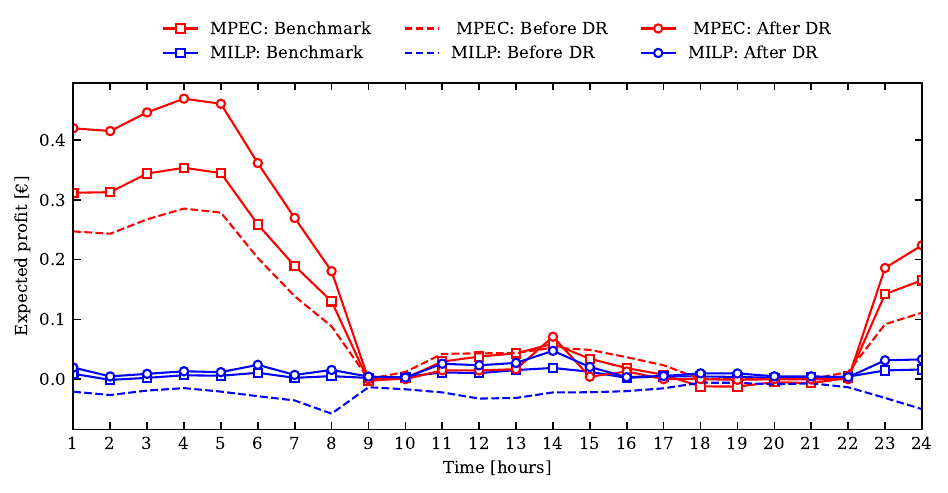}
 \caption{Comparison of the expected profit between the two models before and after DR implementation. }
 \label{P_3_3}%
\end{figure}
The expected profits of the retailer are depicted in \cref{P_3_3} for both models. They are directly related to price tariffs and quantities. If the retailer is myopic regarding the market, it loses money both overall and in some periods, as depicted by the MILP result. Note that in such models, the retailer maximizes its expected profit over an entire 24-hour time horizon, focusing more on the overall or cumulative profit than on profits from individual periods.\footnote{The underlying assumption here is that consumers commit to a tariff agreement with the retailer, which is expected to remain valid for the specified time horizon. This approach is both reasonable and practical in the context of current electricity markets, as discussed in \citep{wu2015demand}.}. However, consumer flexibility with DR plays a pivotal role, in enhancing the retailer's expected profits and mitigating financial losses. Notably, in the MPEC models, where the retailer is a strategic player, its expected profits are significantly greater due to its market power in price formation. 

\subsubsection{Market outcomes: Social welfare analysis}

Social welfare (SW) is calculated as follows:
\begin{align}
SW &= \underbrace{\sum_{\omega\in \Omega}\sigma_{\omega}\left(\sum_{j\in J,t\in T}P_tq_{jt\omega}-\sum_{t\in T}(P_{t\omega}^Sq_{t\omega}^S+Cy_{t\omega})\right)}_{\text{Retailer's expected profit}} +\notag \\
&\underbrace{\quad +\sum_{t\in T}\left(P_tq_{jt\omega}-A_{jt\omega}(q_{jt\omega}+\Delta_{jt\omega}^C)+\frac{1}{2}B_{jt\omega}(q_{jt\omega}+\Delta_{jt\omega}^C)^2\right)}_{\text{consumers' benefit minus cost of consumption}} \notag.
\end{align}
  \begin{figure}[ht]
 \centering
 \subfloat[Social welfare with a quadratic utility function.]{\includegraphics[scale=.580]{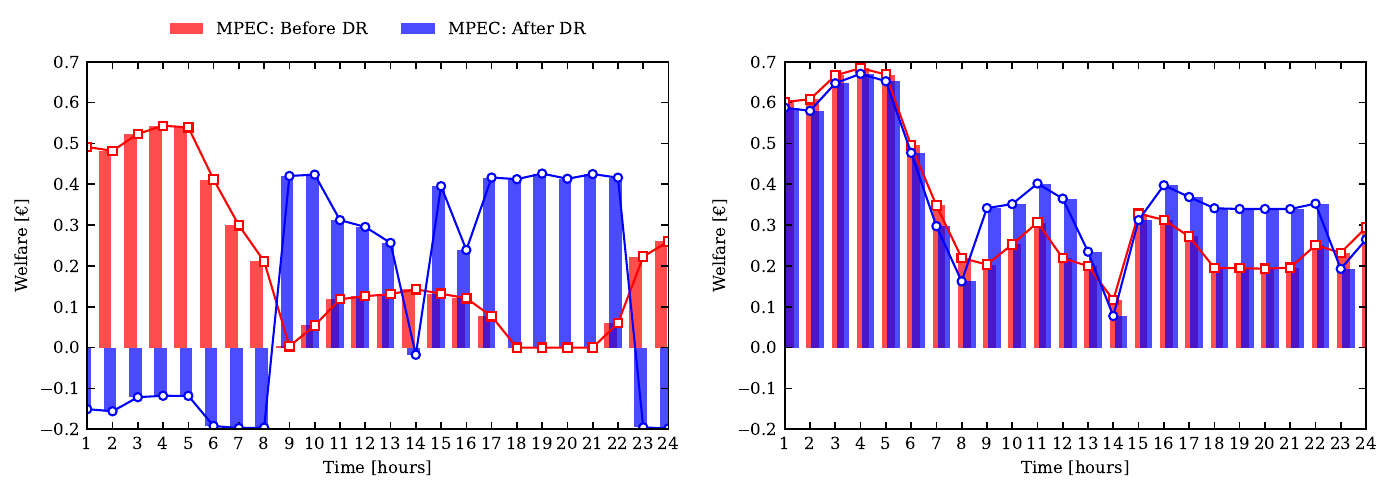}}\\
 \subfloat[Social welfare with a linear utility function.]{\includegraphics[scale=.580]{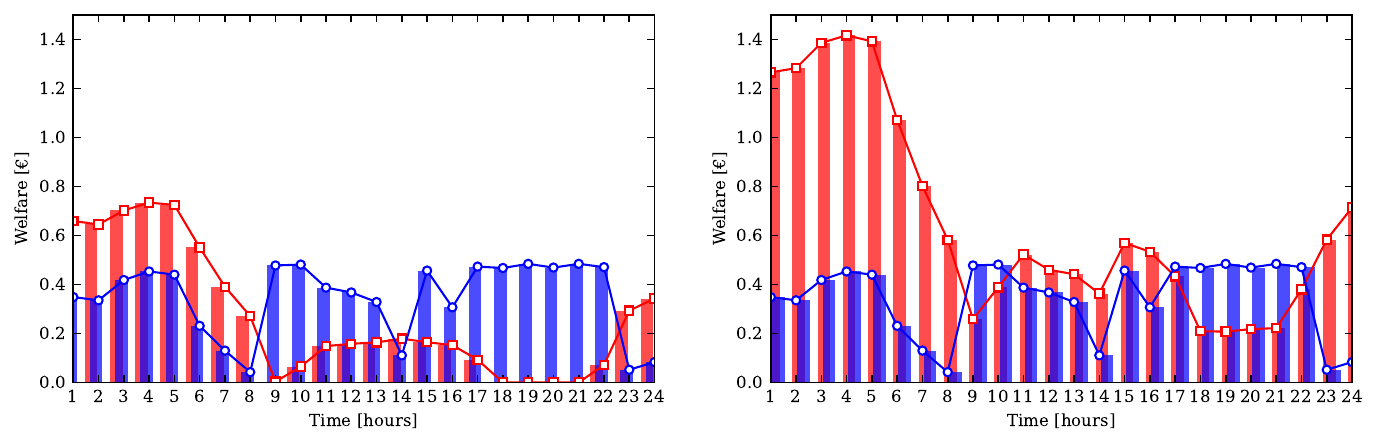}}
 \caption{Social welfare for the MPEC and equilibrium models in the \emph{no flexibility} and \emph{flexibility} cases.}
 \label{CW}%
\end{figure}

\cref{CW} depicts social welfare with two models before and after the implementation of the DR program. The MPEC model shows that the social cost of consumption shifts to low-demand periods using consumers’ flexibility. This shows that retailers with market power can distort consumer flexibility, exploiting it to maximize their expected profits. In doing so, retailers with market power can transfer their risks to consumers. On the other hand, social welfare during peak-demand periods increases as consumers can adjust their consumption during these periods using dynamic price tariffs. Social welfare is better off during low-demand periods due to low-price tariffs, and worse off during peak-demand periods due to high-price tariffs before DR in both models. Consumer flexibility plays an important role in enhancing social welfare from 9:00 to 22:00. These hours are typically stressful for system operators, underlining the significance of consumer flexibility for system stability. Interestingly, the equilibrium model exhibits positive social welfare in all periods. Welfare is greater during peak-demand periods, and the social cost during low-demand periods is significantly lower than that of the MPEC model. 
\cref{CW}(b) presents social welfare before and after implementing the demand response program, using a \emph{linear utility function}. The results from both models demonstrate that social welfare significantly improves with a linear utility function compared to a quadratic one.  During off-peak periods, social welfare is better without consumer flexibility. However, despite differences in the magnitude of welfare, the trend remains consistent whether linear or quadratic utility functions are used. This consistency enables the generalization of market outcomes across the models, regardless of the specific functional form of the consumers' utility functions.
   \begin{figure}
 \centering
  \subfloat[Consumers' utility for each consumer: MPEC.]{\includegraphics[scale=.40]{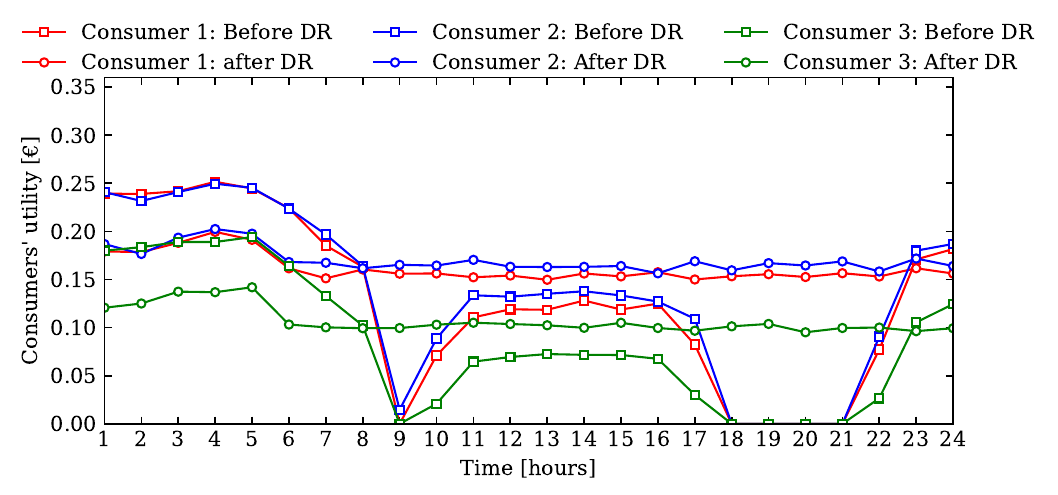}}
   \subfloat[Consumers' welfare for each consumer: MPEC.]{\includegraphics[scale=.40]{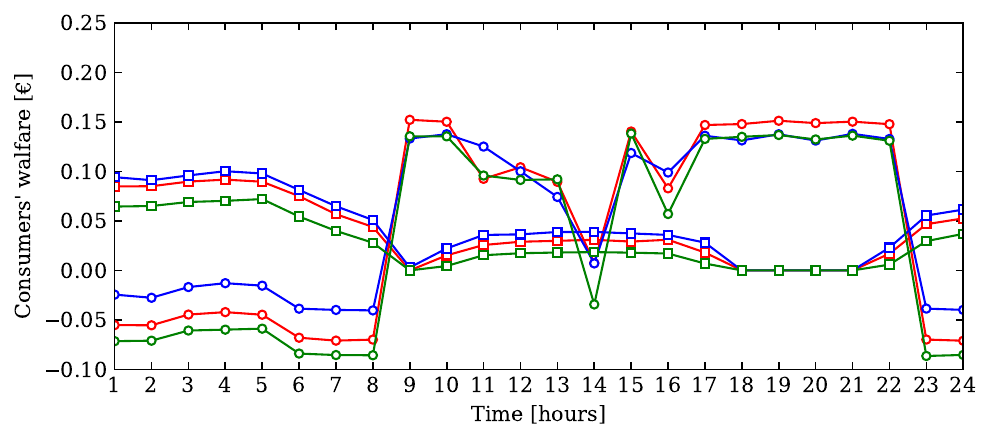}}\\
 \subfloat[Consumers' utility for each consumer: MILP.]{\includegraphics[scale=.40]{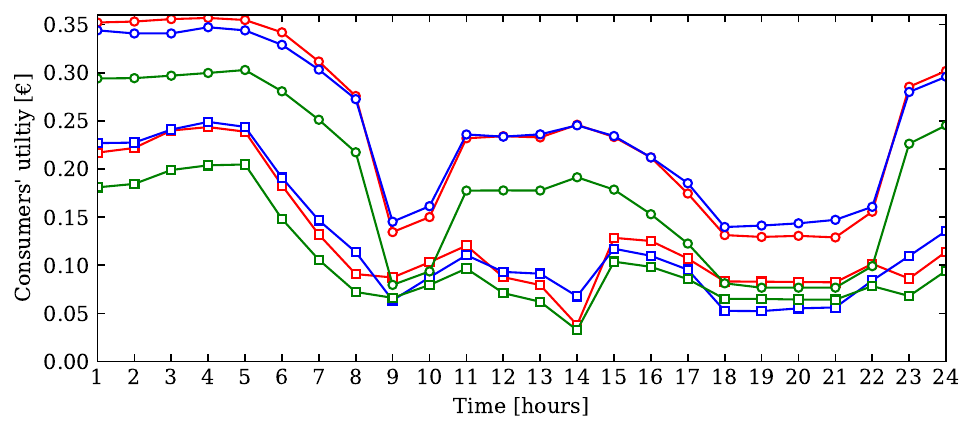}}
 \subfloat[Consumers' welfare for each consumer: MILP.]{\includegraphics[scale=.40]{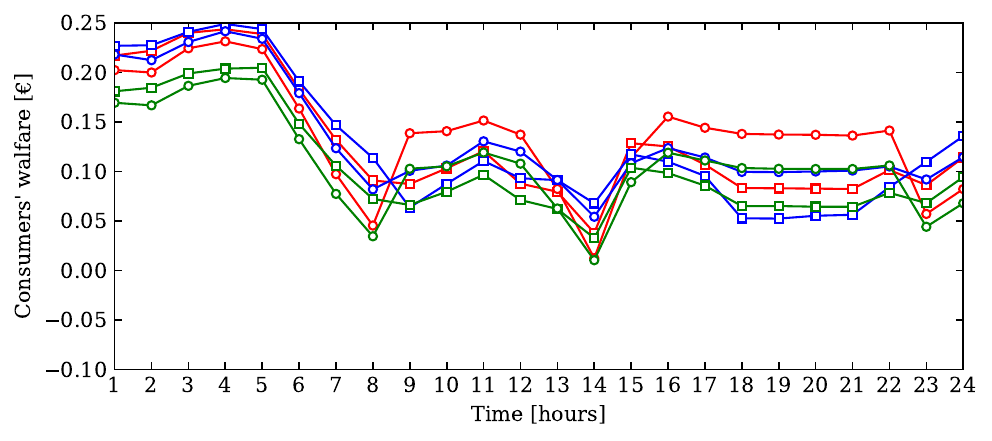}}
 \caption{Comparison of individual consumers' utility and welfare in the MPEC and MILP using \emph{no flexibility} and \emph{flexibility} cases.}
 \label{PP_5}
\end{figure}

\cref{PP_5} shows individual consumers’ welfare in both models, before and after implementing the DR program. Consumers generally experience surpluses, especially during low-demand periods when using MILP. Consumer flexibility leads to a surplus for consumers across all periods compared to their utility without the DR program. Even with the MPEC model, consumers offset their losses during low demand in the periods from 8:00 to 22:00. Note that the differences in individual consumer surpluses are attributable to variations in the utility and demand flexibility parameters, as detailed in \cref{Tab_1}. 

\begin{figure}[H]%
 \centering
 \subfloat[Level of flexibility for consumers: \emph{benchmark}.]{\includegraphics[scale=.50]{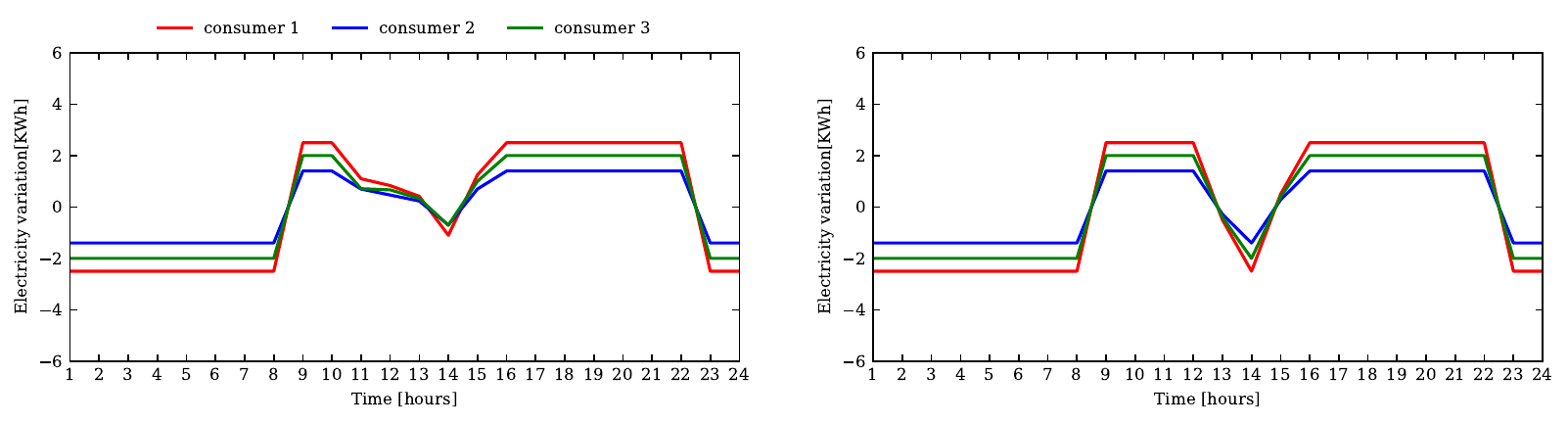}}\\
 \subfloat[Level of flexibility for consumers: \emph{flexibility}.]{\includegraphics[scale=.50]{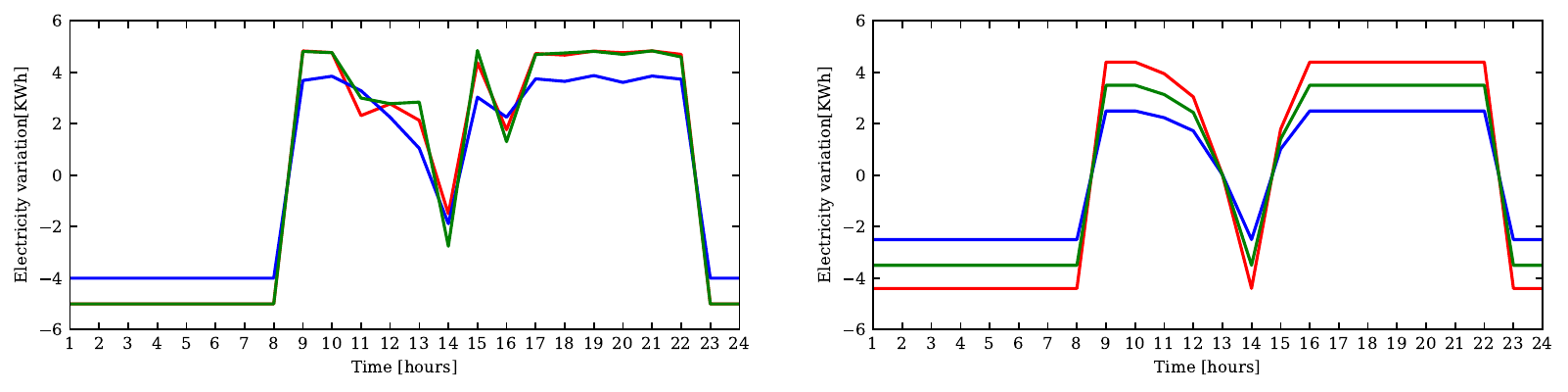}}\\
 \caption{Variation in demand in both models (left MPEC and right MILP) concerning the benchmark and flexibility cases.}
 \label{P_3_8}%
\end{figure}

\cref{P_3_8} presents consumer flexibility in the MPEC and MILP models for the \emph{benchmark} and \emph{flexibility} cases. The overall trend for flexibility is similar in both models. Notably, consumers demonstrate greater flexibility from 9:00-11:00 and 17:00-22:00 by shifting their consumption to off-peak periods. The results adhere to the constraint $\sum_{t\in {T}}\Delta_{jt\omega}^{C}=0$ and respect the flexibility boundary values for each consumer as presented in \cref{Tab_1}. For example, consumer 1 exhibits greater flexibility in consumption, up to 5 kWh, compared to other consumers. With high flexibility, the overall electricity price decreases, and the quantity purchased increases, leading to robust impacts on both the retailer's expected profit and consumer surplus. Therefore, the DR models effectively characterize market outcomes under both market configurations. They highlight the importance of utility functions in understanding consumer behavior and market dynamics.
\subsubsection{Market performance and sensitivity analysis}

 \begin{table}
 \caption{Market performance of the retailer and consumers in the simulations with the \emph{benchmark} case. All values, except for the price (\euro/kWh), represent average outcomes for the considered scenarios ($\Omega = 30$) and are expressed in \euro.}
 \resizebox{\textwidth}{!}{%
\centering
\begin{tabular}{lll|lll}
\hline
Retailer& \multicolumn{2}{c|}{Model} & \multicolumn{1}{c}{Consumers} & \multicolumn{2}{c}{Model}\\ \cline{2-3} \cline{5-6} 
Performance index & MPEC& MILP & performance index & MPEC& MILP\\ \hline
Expected profit  & 0.1149 & -0.0076 & Cost & 0.2602  & 0.2894\\
Revenue & 0.2602& 0.2894   & Price & 0.0230& 0.0177\\
Cost& 0.1454 & 0.2970 & Welfare & 0.0177 & 0.0784\\ \hline
\end{tabular}}
\label{Tab_4}
\end{table}
\cref{Tab_4} presents the average performance of the two models for both the retailer and consumers under the \emph{benchmark} case. On average, the retailer incurs losses in the MILP model. The retailer’s costs come from two sources: \textit{i}) the cost of purchasing in the spot market, which assumes perfect information about future consumption; and \textit{ii}) imbalance penalties, which arise due to imperfect information. It is essential to understand that the revenue generated by the retailer translates into costs for consumers, which significantly affects the overall dynamics of the market.
\begin{table}
\centering
\caption{Sensitivity analysis of spot price uncertainty in the MPEC model: \emph{benchmark} case.}
\resizebox{\textwidth}{!}{%
\begin{tabular}{lllllllll}
\hline
Spot price uncertainty & Exp. profit & Max. profit & Min. profit & $q_{t\omega} $ & $q_{jt\omega}$ & Tariff & Utility & Welfare\\ \hline
0.015     & 0.1143 & 0.3555  & -0.0128 & 12.29 & 4.198 & 0.0230 & 0.1046  & 0.0177\\
0.030     & 0.1156 & 0.3551  & -0.0140 & 12.30 & 4.101 & 0.0230 & 0.1046  & 0.0177\\
0.035     & 0.1156 & 0.3547  & -0.0141 & 12.31 & 4.105 & 0.0230 & 0.1047  & 0.0178\\ \hline
\end{tabular}%
}
\label{Tab_sens}
\end{table}
In \cref{Tab_sens}, we present the market outcomes regarding the uncertainty of spot market prices $P_{t\omega}^S$ under the \emph{benchmark} case and in the MPEC model. In contrast, consumers enjoy better utility due to low price tariffs. The results indicate that uncertainties in spot market prices do not significantly influence market outcomes for either retailers or consumers. As demonstrated by our simulation results, the most crucial parameter affecting market outcomes is consumer flexibility. Thus, consumer flexibility is key not only in mitigating the potential impact of the retailer's market power on the market and in influencing spot market price formation. This highlights the pivotal role of consumer flexibility in managing market dynamics and aiding the integration of renewable energy sources.

 \section{Conclusions}\label{se_P_3_3}
 
This paper introduces two game-theoretical models to analyze retailer-consumer interactions in electricity markets under dynamic tariff-based demand response programs. It models consumer flexibility under dynamic pricing and employs a utility function to characterize consumer behavior, providing insights into market dynamics and reshaping their interpretation. 

Two market structures are considered: a leader-follower model using a bilevel problem solved as MPEC and a perfect competition model solved as a MILP problem. The MPEC model facilitates a comparison of \textit{market power} dynamics, while the MILP model helps in understanding competitive market behavior. Both models reveal that the impact of consumer flexibility on market outcomes and social welfare depends on the specific market configuration. The paper advances existing research on demand response programs under dynamic price tariffs by incorporating uncertainty and levels of competition. It also explicitly accounts for consumer preferences through a utility function. These dynamic interactions allow for a more complete understanding of market dynamics and the potential of consumer flexibility. They further provide managerial insights for regulators on how to leverage this potential. 

The results demonstrate the effectiveness of dynamic pricing in demand response programs for optimizing strategic retailers' expected profit and maximizing consumers' welfare. It also presents economic incentives in the form of price reductions for flexible consumers under price uncertainties, which benefits both retailers and consumers.

From the results, we can draw the following managerial and economic insights: \textit{i}) the proposed models manage consumers' demand and price uncertainties with stochastic programming, \textit{ii}) the results show that market configurations play an important role in the effectiveness of consumer flexibility in demand response based on dynamic tariffs, \textit{iii}) it is demonstrated that consumer flexibility could be crucial in devising market strategies that are attuned to individual needs and adapt to varying market conditions, and \textit{iv}) the results reveal that utilizing consumer flexibility under tariff-based demand response renders significant cost savings and economic efficiency in day-ahead and retailer electricity markets. The takeaway is that enhancing consumer flexibility through demand response could play a crucial role in the sustainability and resilience of future electricity markets.

The proposed models can be extended in different directions.  \textit{i}) Including the operational dynamics of generating companies' electricity generators and market operators might provide further insights into how to exploit the potential for consumer flexibility. 
\textit{ii}) The models could be extended by explicitly accounting for global renewable requirements in operational decision-making and low-level smart grid constraints to facilitate the inclusion of prosumers in such market dynamics.
\textit{iii}) In light of the scant research on the environmental ramifications of demand response programs, integrating environmental metrics into demand response models is possible. This could provide robust policy guidance for harmonizing technological advancements, market mechanisms, and innovative approaches within electricity markets. 

	\section*{Acknowledgements}
 Arega Getaneh Abate has received funding from the European Union’s Horizon $2020$ research and innovation program under the Marie Skłodowska-Curie grant agreement No $899987$. Carlos Ruiz gratefully acknowledges the financial support from the Spanish government through projects PID2020-116694GB-I00 and from the Madrid Government (Comunidad de Madrid) under the Multiannual Agreement with UC3M in the line of “Fostering Young Doctors Research” (ZEROGASPAIN-CMUC3M) and in the context of the V PRICIT (Regional Programme of Research and Technological Innovation. 
\bibliographystyle{elsarticle-num-names}
		\newpage
		\section*{References}
	\bibliography{Reference-WP.bib}
\end{document}